\documentclass[%
reprint,
 amsmath,amssymb,
 aps,superscriptaddress,
]{revtex4-2}

\usepackage{graphicx}
\usepackage{dcolumn}
\usepackage{bm}
\usepackage[mathlines]{lineno}

\usepackage{hyperref}
\usepackage{epsfig}
\usepackage{color}
\usepackage{float}
\usepackage[normalem]{ulem}
\usepackage{url}
\usepackage[margin=1cm]{geometry}
\usepackage{multirow}
\usepackage{array}
\usepackage{booktabs}
\usepackage{xcolor}
\usepackage{amssymb}

\begin{document}

\preprint{APS/123-QED}

\title{\textbf{Single-particle potentials in asymmetric nuclear matter within the LOCV framework} 
}%

\author{Z. Ziarati}
\email{Ziaratizahra@ut.ac.ir }
\address{Department of Physics, University of Tehran, Tehran 14395-547, Iran}

\author{H. R. Moshfegh}
\email{hmoshfegh@ut.ac.ir }
\address{Department of Physics, University of Tehran, Tehran 14395-547, Iran}

\address{Centro Brasileiro de Pesquisas Fısicas, Rua Dr. Xavier Sigaud,
150, URCA, Rio de Janeiro CEP 22290-180, RJ, Brazil}

\date{\today}
\begin{abstract}

The single-particle potentials and effective masses of protons and neutrons in asymmetric nuclear matter are investigated within the lowest-order constrained variational (LOCV) method. The dependence of these quantities on momentum, density, and isospin asymmetry is studied using the $Reid68$ and $AV_{18}$ nucleon-nucleon interactions, with the effects of three-body forces also examined. The neutron and proton single-particle potentials exhibit a systematic splitting with increasing asymmetry, while their decomposition into different interaction channels identifies the dominant contributions to the in-medium interaction. The symmetry potential decreases with increasing momentum and shows a pronounced dependence on density and three-body forces, particularly at low momentum. The neutron-proton effective-mass splitting increases with asymmetry, with the neutron effective mass larger than the proton effective mass. The calculated symmetry potential is also compared with results from other microscopic and phenomenological approaches and with the empirically constrained Lane potential, showing a consistent decreasing trend with increasing kinetic energy. These results provide a microscopic, state-dependent single-particle potential for asymmetric nuclear matter, with its momentum, density, and asymmetry dependence determined directly from the underlying two and three-body interactions.

\end{abstract}

     
\maketitle

\section{\label{sec:level1}Introduction:}

 Nuclear matter provides a fundamental theoretical framework for investigating the properties of strongly interacting many-body systems. The study of nuclear matter remains one of the central topics in nuclear physics owing to its importance for both finite nuclei and dense astrophysical systems. The investigation of its structure and properties is not only of considerable theoretical interest but also essential for understanding a wide variety of nuclear and astrophysical phenomena.

From stable and unstable nuclei \cite{HAGINO_2013} to heavy–ion collisions \cite{Russotto:2023ari}, and from phase transitions in 
dense matter to the internal composition of neutron stars, numerous phenomena are closely related to accurate description of nuclear matter properties. Therefore, nuclear matter provides an ideal theoretical framework for studying effective interactions among nucleons under various physical conditions such as density, temperature, and momentum.

In the simplest case, symmetric nuclear matter, in which the proton and neutron densities are equal, serves as a useful benchmark for exploring the fundamental features of nuclear interactions. However, many realistic physical systems, including finite nuclei and, in particular, neutron stars, exhibit significant isospin asymmetry. Therefore, considerable attention has been devoted to the study of asymmetric nuclear matter and its equation of state. In this context, quantities such as the nuclear symmetry energy, the symmetry potential and the neutron-skin
thickness are of particular importance, since they provide valuable information on the isospin dependence of nuclear
interactions and the properties of neutron-rich matter \cite{Li_1998,DANIELEWICZ2003233,LATTIMER2007109,STEINER2005325,BARAN2005335,LI2008113,PhysRevC.85.024305}.

Among the various properties of nuclear matter, the single-particle
potential (SPP) is of particular importance, as it plays a fundamental
role in determining nucleonic dynamics and the properties of nuclear
systems. The SPP constitutes one of the basic ingredients in microscopic
many-body descriptions of nuclear matter and provides essential
information on the in-medium behavior of nucleons. In particular, an
accurate knowledge of the single-particle potential is crucial for
simulations of the dynamical evolution of heavy-ion collisions at
intermediate and high energies \cite{BERTSCH1988189,AICHELIN1991233}.

It has also been shown that the momentum dependence of the nuclear
single-particle potential must be properly incorporated in theoretical
models in order to reproduce experimental observations
\cite{BERTSCH1988189,PhysRevLett.58.1926,AICHELIN1991233}. Since the SPP embodies the
effects of the surrounding medium on the propagation of nucleons, it
provides valuable information on many-body correlations and in-medium
modifications of the nucleon-nucleon interaction. Furthermore, the real
part of the nucleon optical potential is closely related to the nuclear
single-particle potential, allowing microscopic calculations to be tested
against experimental information from nucleon-nucleus scattering and
heavy-ion collisions.

 Moreover, the momentum and isospin dependence of the SPP are intimately
connected to several key properties of asymmetric nuclear matter. The
momentum dependence of the single-particle potential determines the
nucleon effective mass, while its isovector component, characterized by
the difference between the neutron and proton potentials, defines the
nuclear symmetry (Lane) potential. The symmetry potential plays a central
role in determining the density dependence of the nuclear symmetry
energy and is directly related to it through the Hugenholtz--Van Hove
theorem \cite{PhysRevC.82.054607,LI2010509c}. As a key input to nuclear
transport models, it governs phenomena such as isospin diffusion and
collective transverse flow in heavy-ion collisions, thereby providing
access to the symmetry-energy slope parameter $L$
\cite{PhysRevLett.94.032701}. The isovector potential can also be
constrained by nucleon-nucleus scattering data, $(p,n)$ charge-exchange
reactions, and single-particle bound-state energies
\cite{PhysRevC.82.054607}, while its density dependence is correlated
with the neutron-skin thickness of heavy nuclei
\cite{PhysRevC.72.064309}.

In asymmetric matter, the neutron-proton effective-mass splitting is
another important manifestation of the isovector momentum dependence of
the SPP and is closely related to the energy dependence of the symmetry
potential \cite{Li_2015}. The sign and magnitude of the splitting effective masses,
$m_n^*-m_p^*$, are important for understanding the momentum dependence
of isospin-asymmetric nuclear dynamics and for constraining transport
calculations of heavy-ion collisions \cite{Li_2015,PhysRevC.82.054607}.

Despite considerable progress, a quantitative description of the single-particle properties of asymmetric nuclear matter remains a challenging problem. Significant uncertainties still persist regarding the density and momentum dependence of neutron and proton single-particle potentials, the neutron–proton effective-mass splitting, and their behavior at supranuclear densities. Furthermore, the microscopic origin of these properties and the relative contributions of different interaction channels are not yet fully understood. In particular, the decomposition of the single-particle potential into various spin-isospin and partial-wave channels provides valuable insight into the mechanisms governing the in-medium interaction and allows one to identify the dominant channels responsible for the isospin dependence of nuclear matter properties. Such channel-by-channel analyses are particularly useful for clarifying the role of tensor correlations and short-range
interactions in asymmetric matter.

Consequently, numerous  theoretical microscopic
approaches have so far been employed to investigate the single–particle potential of nuclear matter, including the Brueckner–Hartree–Fock (BHF) method \cite{PhysRevC.40.R491,INSOLIA199412,ZUO19981,PhysRevC.59.2927,ZUO2010574c,PhysRevC.74.014317,PhysRevC.73.035208,PhysRevC.72.034005}, the Dirac–Brueckner–Hartree–Fock (DBHF) approach \cite{Sammarruca_2005,PhysRevLett.56.1237,PhysRevC.48.2707,LEE1997235,PhysRevC.72.065803,PhysRevC.76.054316,Sammarruca_2010}, the medium T-matrix (MTM) and Green’s function (GF) methods \cite{RAMOS19891,PhysRevC.78.054003,PhysRevLett.90.152501,RIOS2007346}, as well as variational Fermi hypernetted-chain (FHNC) calculations \cite{FRIEDMAN1981205}.

Among the various microscopic many-body approaches, the lowest-order constrained variational (LOCV) method has proven to be a reliable and efficient framework for investigating the properties of strongly interacting many-body systems. The LOCV method was developed in the late 1970s and subsequently been applied successfully to other many-body systems\cite{MODARRES2023104047}.

One of the main advantages of the LOCV approach is its fully microscopic and self-consistent nature. In contrast to phenomenological models, the method does not introduce additional adjustable parameters beyond those contained in the underlying two- and three-body interactions employed as input. Furthermore, the cluster expansion exhibits rapid convergence due to the constrained variational treatment of short-range correlations. The formalism naturally incorporates channel-dependent correlation functions, allowing one to investigate separately the contributions of different spin-isospin and partial-wave channels to various properties of nuclear matter.

 Another important feature of the LOCV method is its
flexibility and broad applicability. The formalism can be
straightforwardly extended to include three-body forces \cite{Goudarzi2015} 
and finite-temperature effects \cite{MOSHFEGH2007201,sepah2003} and has been successfully
applied to symmetric and asymmetric nuclear matter \cite{zar2010},
hyperonic matter \cite{SHAHRBAF201966, PhysRevC.100.044314}, neutron-star matter \cite{Khanmo} over the past three decades. Owing to these advantages, the LOCV approach has become a well-established microscopic tool for studying the equation of state and single-particle properties of nuclear systems.

Previous LOCV investigations of single-particle properties
have mainly been restricted to symmetric nuclear matter
and have considered only two-body interactions \cite{MODARRES20111}.
Consequently, the effects of isospin asymmetry and
three-body forces on neutron and proton single-particle
potentials remain largely unexplored within the LOCV
framework. Moreover, a systematic channel decomposition
of the single-particle potential in asymmetric nuclear
matter has not yet been performed.

Motivated by the microscopic and channel-dependent
nature of the LOCV formalism, in the present work we
extend previous studies to asymmetric nuclear matter and
perform a systematic investigation of neutron and proton
single-particle potentials. Particular attention is devoted
to their density, momentum, and asymmetry dependence,
as well as to the role of three-body forces and the
contributions of different spin-isospin and partial-wave
channels.

The paper is organized as follows: Section \ref{sec:Formalism} provides a brief explanation of the LOCV method and  SPP calculation for asymmetric nuclear matter. Section \ref{sec:result and discussion} presents the numerical results and analyzes their dependence on density, momentum, and the asymmetry parameter. Finally, Section \ref{sec:conclusion} summarizes the main conclusions and
presents some perspectives for future investigations. 
\section{theoretical framework}
\label{sec:Formalism}

\subsection{LOCV description of asymmetric nuclear matter}

The properties of asymmetric nuclear matter are investigated within the framework of the lowest-order constrained variational (LOCV) method. The Hamiltonian of the system is written as

\begin{equation}
{H}=\sum_i T_i
+\sum_{i<j}V_{ij}
+\sum_{i< j< k}V_{ijk},
\end{equation}
where $T_i$, $V_{ij}$, and $V_{ijk}$ denote the kinetic-energy operator, the two-body  and the three-body interaction, respectively. The present formalism is independent of the particular choice of nuclear interactions and can therefore be applied to any realistic two- and three-body potentials.

The isospin asymmetry of nuclear matter is characterized by the asymmetry parameter

\begin{equation}
\beta=\frac{\rho_n-\rho_p}{\rho},
\qquad
\rho=\rho_n+\rho_p,
\end{equation}

where $\rho_n$ and $\rho_p$ are the neutron and proton densities, respectively. The limiting cases $\beta=0(1)$  correspond to symmetric nuclear(pure neutron)matter.

In the LOCV approach, the variational trial wave function is chosen as,
\begin{equation}
\Psi =\mathcal{F}\Phi,
\end{equation}

where $\Phi$ is the Slater determinant of noninteracting nucleons, i.e. plane waves,

\begin{equation}
\Phi = \mathcal{A} \prod_{i} \exp(i\mathbf{k}_i \cdot \mathbf{r}_i),
\end{equation}
and $\mathcal{F}$ is the $N$-body correlation operator, which is approximated by the Jastrow type, i.e.,:
\begin{equation}
\mathcal{F}(1\cdots N) = \mathcal{S} \prod_{i>j} f(ij),
\end{equation}

with $\mathcal{A}$ and $\mathcal{S}$ denoting the antisymmetrization and symmetrization operators, respectively.

The energy per particle is obtained through the cluster expansion,
\begin{equation}
E=\frac{1}{N}
\frac{\langle\Psi|H|\Psi\rangle}
{\langle\Psi|\Psi\rangle}=E_1+E_2+\cdots ,
\end{equation}

where $E_1$ and $E_2$ are the one-body and two-body contributions, respectively. The normalization constraint is introduced in such a
way that the contribution of higher-order clusters is
minimized through an appropriate choice of the
correlation functions. Consequently, the neglected
three- and higher-body cluster energies remain small,
thereby justifying the truncation of the cluster
expansion at the two-body level
\cite{HRMoshfegh_1998,OWEN1976170}.

The one-body term corresponds to the kinetic energy of an asymmetric free Fermi gas,

\begin{equation}
E_1=
\sum_{\tau =n,p}
\frac{3}{5}
\frac{\hbar^2 {k_F^{\tau}}^2}{2m_{\tau}}
\frac{\rho_{\tau}}{\rho},
\end{equation}

where $k_F^{\tau}=(3\pi^2\rho_\tau)^{1/3}$, $m_\tau$ and $\rho_ \tau$ denote the Fermi momentum, mass and number density of particle species $\tau$, respectively. $\rho$ is the total baryon number density.

The two-body energy contribution is given by
\begin{equation}
\label{eq:8}
E_2=
\frac{1}{2N}
\sum_{ij}
\langle ij|
\mathcal{V}(12)
|ij\rangle_a ,
\end{equation}
where the two-particle states are written as

\begin{equation}
\lvert ij \rangle = \lvert \mathbf{k}_i, \sigma_i, \tau_i ; \mathbf{k}_j, \sigma_j, \tau_j \rangle,
\end{equation}
where $\mathbf{k}$, $\sigma$, and $\tau$ denote the momentum, spin, and isospin quantum numbers of the nucleons, respectively. The subscript $a$ denotes anti-symmetrization. The effective two-body interaction operator is defined as: 

\begin{equation}
\label{eq:10}
\mathcal{V}(12)=-\frac{\hbar^2}{2m}
\left[
f(12),
\left[\nabla_{12}^{2},f(12)\right]
\right]
+
f(12)V(12)f(12).
\end{equation}

The two-body correlation operator is expanded as
\begin{equation}
f(12)=\sum_{\alpha,p}
f_{\alpha}^{(p)}(r_{ij})
O_{\alpha}^{(p)}(ij),
\end{equation}

where $\alpha={J,L,S,T,M_T}$ denotes the two-body quantum numbers, with $J, L, S, T,$ and $M_T$ being the total angular momentum, orbital angular momentum, total spin, total isospin, and the third component of isospin, respectively. The operators $O_{\alpha}^{(p)}$ distinguish between coupled and uncoupled channels.

The correlation functions are obtained by minimizing the two-body energy functional under the normalization constraint
\begin{equation}
\label{eq:12}
\frac{1}{N}
\sum_{ij}
\langle ij|
{f^{P}_{M_T}}^{2}(12)
-f^{2}(12)
|ij\rangle_a
=0,
\end{equation}

where $f^{P}_{M_T}(12)$ denotes the modified Pauli function. For asymmetric nuclear matter the modified Pauli function is given by:

\begin{equation}
f^{P}_{M_T}(r)=
\begin{cases}
\left[1-\dfrac{9}{2}\left(\dfrac{J_1(k_F r)}{k_F r}\right)^2\right]^{-1/2}, & M_T=\pm 1,\\[6pt]
1, & M_T=0.
\end{cases}
\end{equation}
The minimization of Eq. \ref{eq:8} under the normalization
constraint of Eq. \ref{eq:12} introduces a Lagrange multiplier
and leads to a set of coupled and uncoupled
Euler-Lagrange differential equations for the
correlation functions \cite{MOSHFEGH200579}. The resulting density- and asymmetry-dependent correlation functions are then employed to construct the effective interaction of the medium. Detailed derivations of the Euler-Lagrange equations
and their numerical implementation can be found in
Refs. \cite{MOSHFEGH200579,MOSHFEGH2007201}.

The contribution of the three-body force is incorporated self-consistently through an effective density-dependent two-body interaction. Following the procedure of Ref.~\cite{Goudarzi2015}, the degrees of freedom associated with the third nucleon are integrated out, where the averaging is weighted by the two-body correlation functions obtained within the LOCV framework. Consequently, the effective interaction and the correlation functions are determined self-consistently at each density and asymmetry.

\subsection{Nucleon single-particle potential in asymmetric nuclear matter}
The energy of a nucleon with momentum $k_i$ is given by
\begin{equation}
e(k_i) = \frac{\hbar^2 k_i^2}{2m} + U(k_i),
\end{equation}
where $U(k_i)$ denotes the single-particle potential (SPP), representing the average interaction of nucleon $i$ with the surrounding medium.
Within the LOCV method, the  SPP can be identified by comparing the total energy functional with the standard expression for the energy of an interacting Fermi system. This leads to
\begin{equation}
U(k_i)=\sum_{j}\,\langle ij \vert \mathcal{V}(12) \vert ij \rangle_{a},
\label{U(ki)}
\end{equation}
 where $\mathcal{V}(12)$ is the effective interaction introduced in Eq.~\ref{eq:10}. The summation extends over all occupied states of the second particle $j$, while the quantum state of particle $i$ remains fixed.

After inserting  complete sets of relative and center-of-mass states and performing the partial-wave decomposition, the single-particle potential can be expressed as
\begin{equation}
\label{U}
\begin{split}
U(k_i) = {} & \frac{4\pi}{\Omega}\sum_{\alpha}\sum_{k_j}\sum_{\tau_j}(2J+1) \left( \frac{1-(-1)^{L+S+T}}{2} \right) \\
& \times \Big| \langle m_{\tau_i}m_{\tau_j}|TM_T\rangle \Big|^2 \int_0^{\infty} r^2\,dr j_l^2(k_{ij}r)\, \mathcal{V}^\alpha(r) ,
\end{split}
\end{equation}
 where $k_{ij} = \frac{1}{2} |\mathbf{k}_i - \mathbf{k}_j|$, and $r = |\mathbf{r}_i - \mathbf{r}_j|$ are the relative momentum and relative distance of the interacting pair.  $\mathcal{V}^\alpha(r) = \langle \alpha | \mathcal{V}(r) | \alpha \rangle$ denotes the channel-dependent effective interaction and $\langle m_{\tau_i}m_{\tau_j}|TM_T\rangle$ are
the Clebsch-Gordan coefficients, where $m_{\tau_i}$ and $m_{\tau_j}$ denote the
third components of the isospin of nucleons
$\tau_i$ and $\tau_j$, respectively.. 

At zero temperature, the occupation probability is described by the step function, $n(k)=\Theta(k_{F_{\tau}} - k_j)$ where $k_{F_{\tau}}$ denotes  the neutron or proton fermi momentum. Consequently, the summation over the momentum states of particle $j$ can be transformed into an integration over the corresponding Fermi sea according to

\begin{equation}
\sum_{{k}_j}\;\to\;\frac{\Omega}{(2\pi)^3}\int d^3k_{j}\, n(k_{j}).
\end{equation}

After carrying out the momentum and angular
integrations, the single-particle potential of a nucleon
with isospin projection $ \tau_i$ can be written as
\begin{equation}
\label{eq:18}
U_{\tau_i}(k_{\tau_i},\rho,\beta)
=
\sum_{\tau_j=n,p}
U_{\tau_i\tau_j},
\end{equation}
where the index $\tau_j$ specifies the species of the interacting nucleon occupying the medium. The quantity $U_{\tau_i\tau_j}$ therefore represents the contribution to the single-particle potential of particle $\tau_i$ arising from its interaction with nucleons of type $\tau_j$. For neutron (proton) single-particle potentials,
the summation in Eq. \ref{eq:18} includes both neutron-neutron
(proton-proton) and neutron-proton interaction channels.

The partial contributions $U_{\tau_i\tau_j}$ are given by:
\begin{equation}
\label{eq:19}
\begin{split}
U_{\tau_i \tau_j}={} & \frac{1}{\pi}\sum_{\alpha}(2J+1) \left( \frac{1-(-1)^{L+S+T}}{2} \right) \\& \times
  \Big| \langle m_{\tau_i} m_{\tau_j}|TM_T\rangle \Big|^2 \int_0^{\infty} r^2\,dr I_{L}(k_{\tau_i} , k_{F} ^{\tau_j}, r)\, \mathcal{V}^\alpha(r),
\end{split}
\end{equation}
where
\begin{equation}
I_{L}(k_i , k_{F}^j , r)=\int_{0}^{k_{F}^j }\int_{0}^{\pi} k_{j}^2\, j_L^2(k_{ij}r)\, \sin\theta \, d\theta \, dk_{j} ,
\label{IL}
\end{equation}
contains the integration over the occupied Fermi sea of particle $\tau_j$. Here,
$j_L(x)$ is the spherical Bessel function of order $L$, and $k_F^{\tau_j}$ is the Fermi momentum of nucleons of type $\tau_j$.

The channel-dependent effective interaction, $\mathcal{V}^\alpha(r)$ 
is obtained from the LOCV formalism discussed in the previous subsection, while the factor$\left(
\frac{1-(-1)^{L+S+T}}{2}
\right)
$
ensures the antisymmetry of the two-nucleon wave function. 

Equations (\ref{eq:18}) and (\ref{eq:19}) constitute the final expressions for the neutron and proton single-particle potentials in asymmetric nuclear matter within the LOCV framework. In the limiting cases of $\beta=0$ and $\beta=1$, the above expressions naturally reduce to those corresponding to symmetric nuclear matter and pure neutron matter, respectively.

The formalism developed above provides a fully
self-consistent framework for investigating the density,
momentum, and isospin dependence of neutron and
proton single-particle potentials in asymmetric nuclear
matter. The resulting mean fields constitute the basis
for the analysis of nucleon effective masses and other
single-particle properties presented in the next section.


\section{result and discussion}
\label{sec:result and discussion}
In this section, we present the neutron and proton single-particle potentials in asymmetric nuclear matter
obtained within the LOCV framework using the $Reid68$ and Argonne $V_{18}$ ($AV_{18}$) interactions. We first analyze the
 momentum, density, and isospin dependence of the single-particle potentials and investigate the role of the three-body force. The microscopic origin of these features is then explored through a channel
decomposition of the effective interaction. Finally, the resulting nuclear symmetry potential and nucleon
effective masses are discussed and compared with other
microscopic many-body calculations.
\subsection{Single-paricle potential}
\label{SPP}
\begin{figure}[htb!]
\centering
\includegraphics[width=\linewidth]{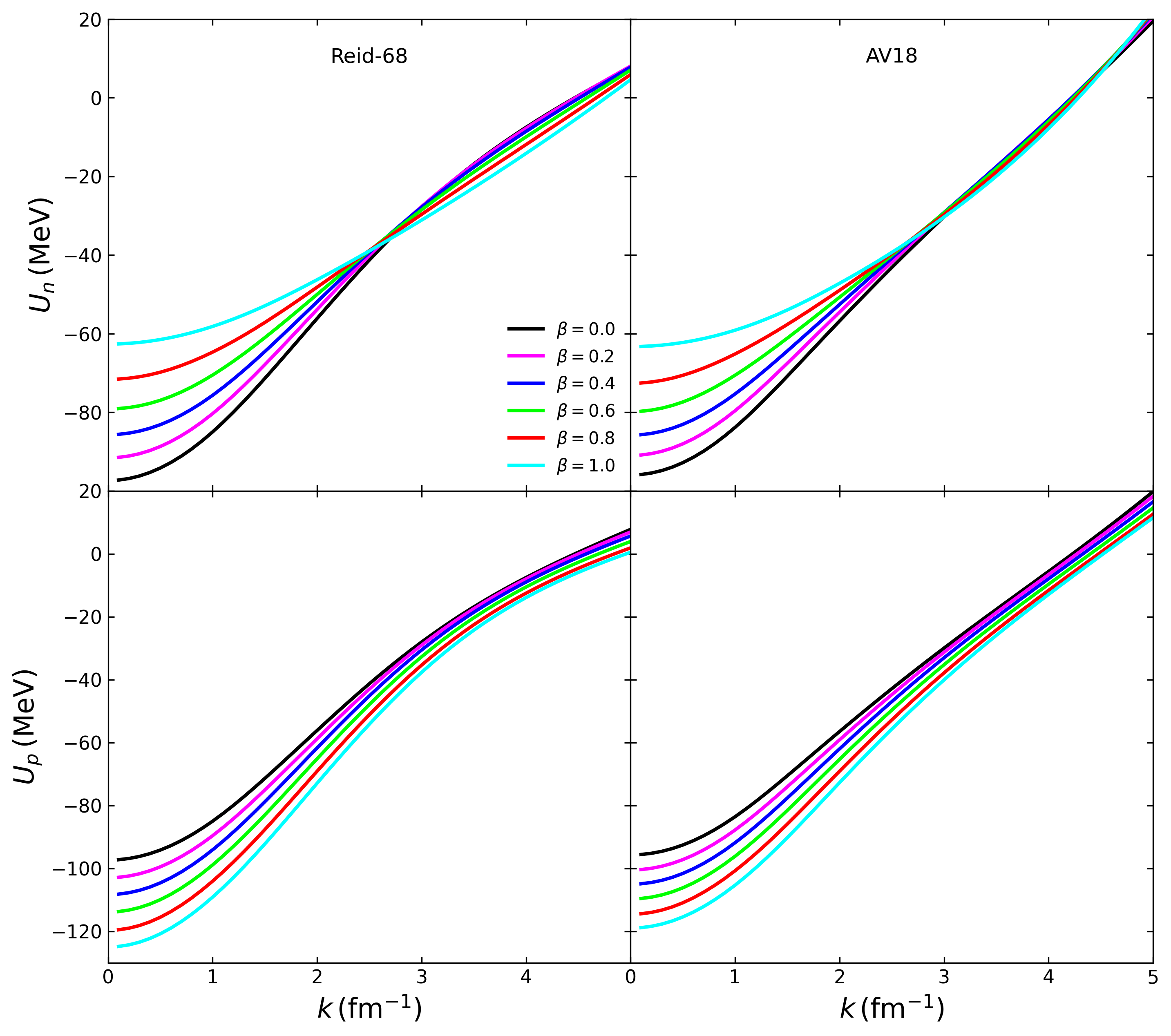}
\caption{\label{fig:1} Momentum dependence of the neutron (upper panel) and proton (lower panel) single-particle potentials at density $\rho=0.17~\mathrm{fm}^{-3}$, calculated with the $Reid68$ (Left Panels) and $AV18$ (right panels) potentials.}
\end{figure}

By solving the LOCV equations with the $Reid68$  and $AV_{18}$ two-body interactions, the correlation functions and, subsequently, the effective interaction in each channel are obtained for each density and isospin-asymmetry parameter ($\beta$). Substitution of the resulting effective interaction into Eq.~\ref{eq:19} then yields the single-particle potential (SPP) as a function of momentum, density, and isospin asymmetry.
Figure \ref{fig:1} displays the neutron and proton SPPs as functions of momentum at a baryon density of $\rho=0.17~\mathrm{fm}^{-3}$ for several values of $\beta$, calculated using the $AV_{18}$ and $Reid68$ interactions. Increasing isospin asymmetry leads to a pronounced splitting between the neutron and proton mean fields. In particular, the proton SPP becomes progressively more attractive with increasing neutron excess, whereas the neutron SPP shifts toward more repulsive values. This behavior reflects the different weighting of like-nucleon and neutron-proton contributions in asymmetric matter. In neutron-rich matter, protons interact predominantly with the more abundant neutrons, while the neutron-proton interaction also contains the isospin-singlet $T=0$ channels, which are absent for identical nucleons. The enhanced contribution of neutron-proton correlations therefore leads to a deeper proton mean field as the neutron excess increases. The calculated isospin dependence is qualitatively consistent with previous microscopic many-body calculations of single-particle potentials in asymmetric nuclear matter \cite{bombaci1991asymmetric,Sammarruca_2005,Dalen_2005}.

The results obtained with the $AV_{18}$  and $Reid68$ interactions exhibit very similar qualitative behavior over the range considered. We therefore use the $AV_{18}$ interaction in the remainder of this work.

\begin{figure}[b]
\centering
\includegraphics[width=\linewidth]{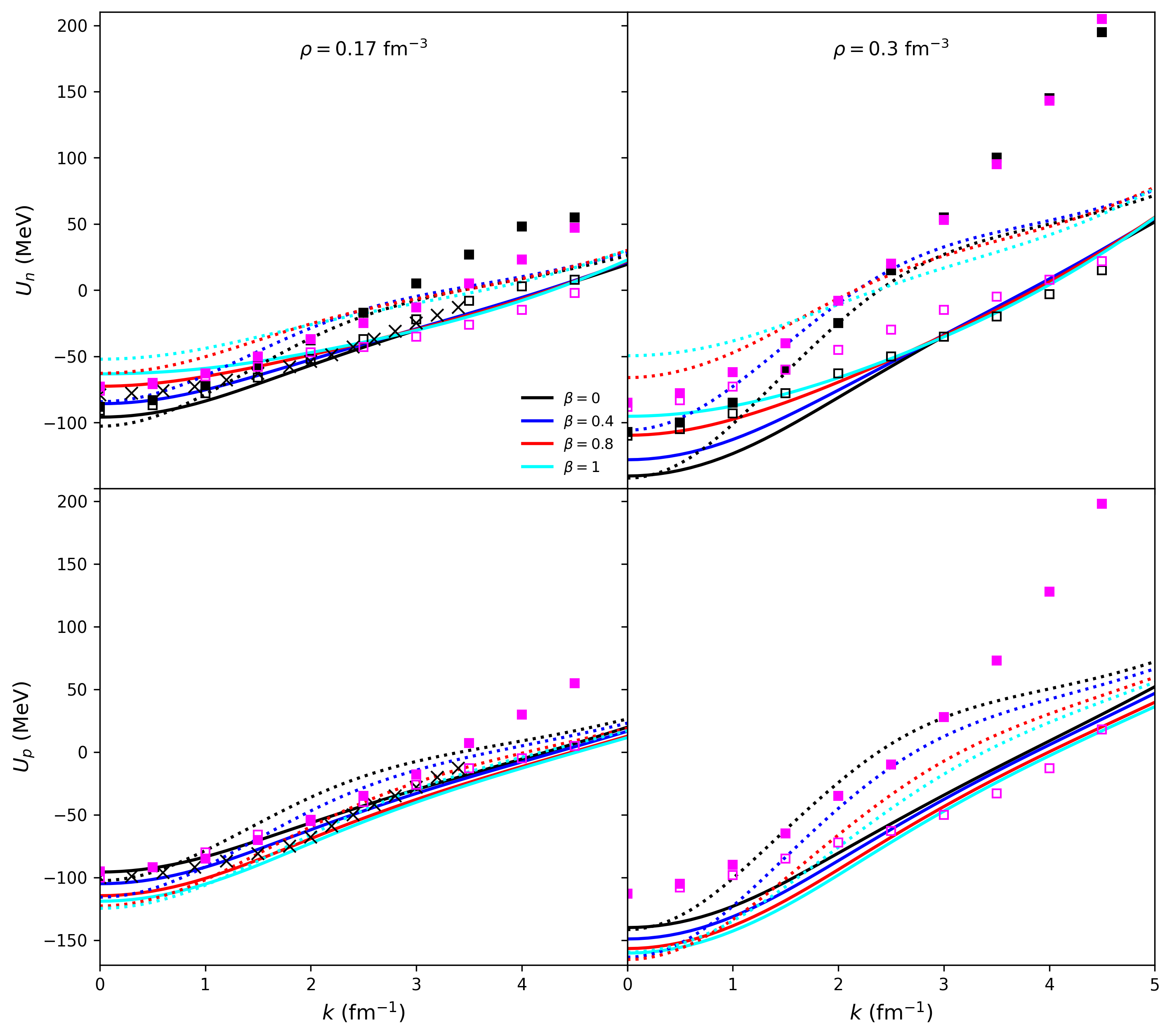}
\caption{\label{fig:2}Neutron (upper panel) and proton (lower panel) single-particle potentials as a function of momentum for two given densities and four isospin asymmetry parameters, $\beta$. The dotted (solid) lines correspond to calculations with (without) the three-body force. Filled (open) squares denote the BHF results obtained with (without) the
three-body force. Black squares correspond to symmetric nuclear matter,
whereas pink squares represent the case of $\beta = 0.4$, respectively \cite{PhysRevC.74.014317}. The BBG results , denoted by $\times$, are also displayed, obtained by considering only the two-body force \cite{bombaci1991asymmetric}. }
\end{figure}

To investigate the effect of the three-body force (TBF) on the nucleon SPPs, Fig. \ref{fig:2} shows their momentum dependence at two baryon densities, $\rho = 0.17$ and $0.3~\mathrm{fm}^{-3}$, for four representative values of the isospin-asymmetry parameter, $\beta=0, 0.4, 0.8$, and $1$. The solid and dotted curves correspond to calculations with the two-body force alone and with the inclusion of the Urbana IX (UIX) three-body force, respectively.

In general, the inclusion of the TBF shifts both neutron and proton SPPs toward more repulsive values. This effect is relatively moderate around saturation density but becomes substantially stronger at $\rho=0.3~\mathrm{fm}^{-3}$, demonstrating the increasing importance of three-body correlations at suprasaturation densities. The magnitude of the TBF contribution also depends on the isospin composition of the system. In particular, the modification of the proton potential becomes increasingly pronounced with increasing asymmetry, while the neutron potential also exhibits a clear repulsive shift. These features reflect the density- and isospin-dependent nature of the three-body interaction and its contribution to the in-medium single-particle field. The repulsive character of the TBF contribution is consistent with microscopic Brueckner calculations, where three-body forces provide an increasingly important repulsive contribution to the single-particle field at high density \cite{PhysRevC.74.014317}.

For comparison, Fig. 2 also includes the neutron and proton SPPs obtained within the BHF approach using the microscopic three-body force of Grangé et al. \cite{PhysRevC.74.014317}, as well as BBG results obtained with the two-body interaction only \cite{bombaci1991asymmetric}. The BHF results with the microscopic TBF are systematically shifted toward more repulsive values relative to the corresponding two-body-force results, particularly at the higher density. This is consistent with the behavior found in the present LOCV calculation. The remaining quantitative differences between the LOCV and BHF results reflect the different treatments of many-body correlations and three-body interactions in the two approaches.

\begin{figure}[H]
\centering
\includegraphics[width=\linewidth]{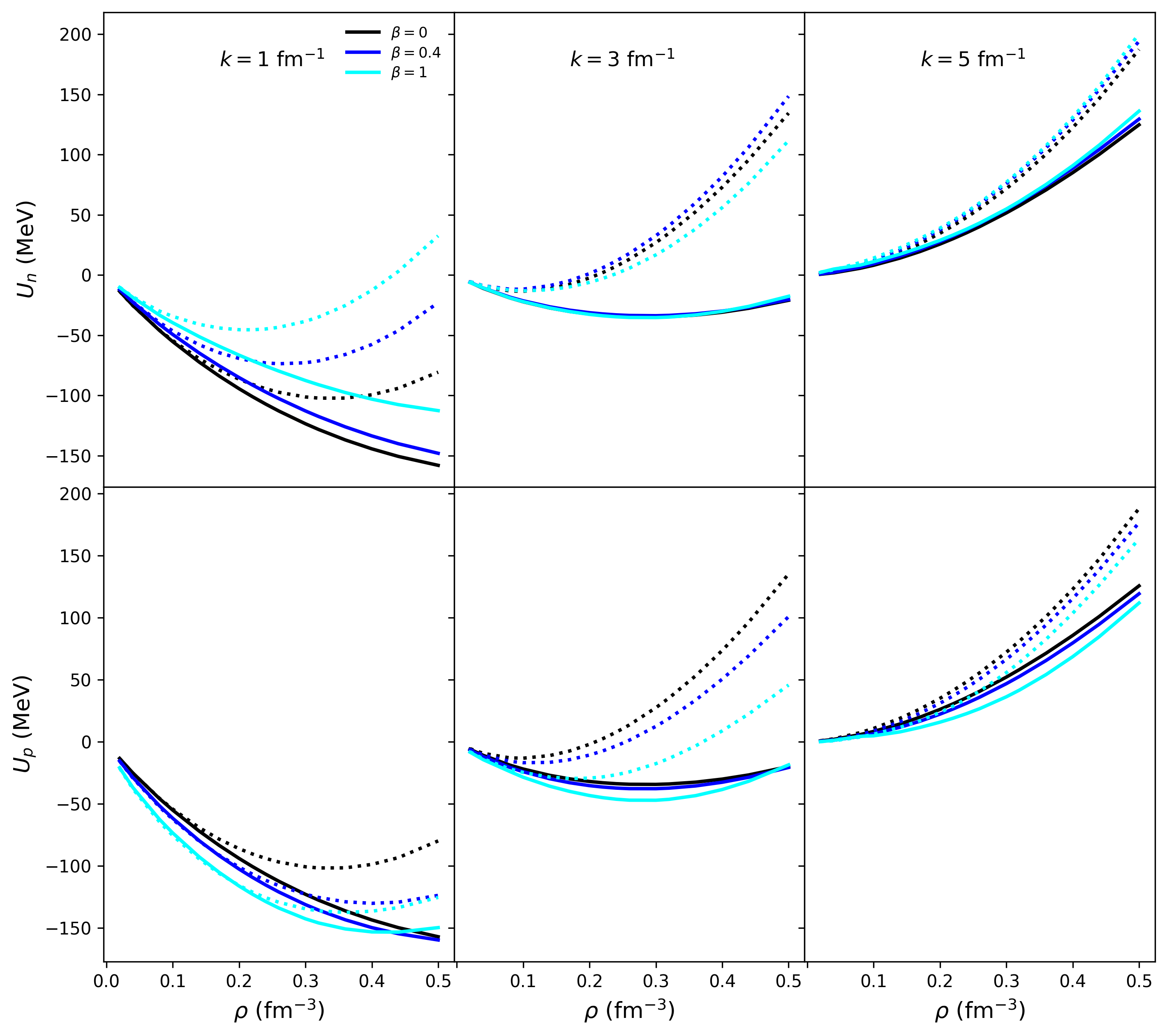}
\caption{\label{fig:3}Density dependence of the neutron single-particle potential (upper panel) and proton single-particle potential (lower panel) at three fixed momenta, calculated using the $AV18$ interaction. The solid curves represent the results obtained with the two-body force only, while the dotted curves show the effect of including the three-body force.}
\end{figure}

The density dependence of the neutron and proton SPPs is examined in more detail in Fig. \ref{fig:3}, where the potentials are shown at three fixed momenta for three representative values of $\beta$, both with and without the TBF. At a given momentum and density, increasing the neutron excess makes the proton SPP more attractive, while the neutron SPP becomes less attractive, reflecting the isovector character of the single-particle field. The momentum dependence is also evident: with increasing momentum, the SPPs become progressively more repulsive, while the neutron-proton splitting becomes less pronounced. At low momentum, the density dependence is nonmonotonic; the potentials initially become more attractive with increasing density, reach a minimum, and then turn toward more repulsive values. This minimum becomes progressively weaker with increasing momentum and essentially disappears at large $k$, where the SPPs increase almost monotonically with density. The inclusion of the TBF modifies this behavior by introducing an additional density-dependent repulsive contribution, which becomes increasingly important at higher densities and is particularly evident at larger momenta. These results demonstrate that the density dependence of the SPP is strongly coupled to its momentum and isospin dependence, with three-body correlations becoming increasingly important away from the low-density, low-momentum regime \cite{PhysRevC.74.014317,Goudarzi2015}.


\subsection{Channel decomposition}
\label{Channel decomposition}
The channel-dependent nature of the LOCV approach allows the single-particle potential to be resolved into contributions from individual partial-wave channels characterized by the quantum numbers $J$, $L$, $S$, $T$, and $M_T$. This decomposition provides a microscopic means of identifying the partial-wave contributions responsible for the momentum and isospin dependence of the neutron and proton single-particle potentials. Unless otherwise stated, all results presented in this subsection include the three-body force.

\begin{figure}[t!]
\centering
\includegraphics[width=\linewidth]{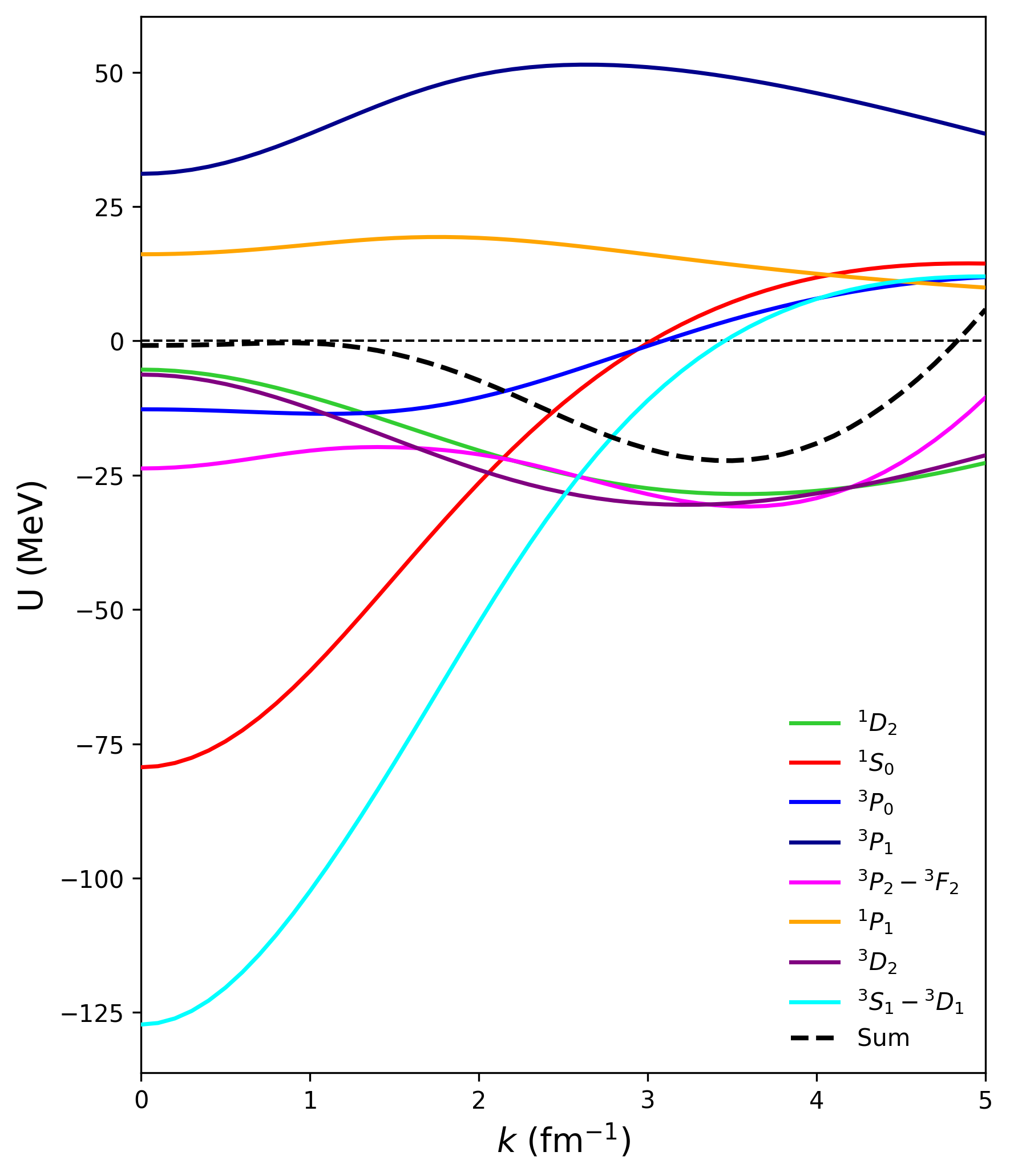}
\caption{\label{fig:4}Contributions of the partial-wave channels with $J \leq 2$ to the single-particle potential $U(k)$ as a function of momentum $k$ in symmetric nuclear matter at $\rho = 0.17~\mathrm{fm}^{-3}$. The dashed curve represents the summation over all contributions excluding the $^3S_1$-$^3D_1$ and $^1S_0$ channels.}
\end{figure}

Figure \ref{fig:4} shows the contributions of the partial-wave channels with $J \leq 2$ to the single-particle potential $U(k)$ in symmetric nuclear matter at density $\rho = 0.17~\mathrm{fm}^{-3}$, as a function of the momentum $k$. Among the individual channels, the $^{3}S_{1}\text{-}^{3}D_{1}$ and $^{1}S_{0}$ channels provide the two largest attractive contributions over the momentum range considered. The $^{3}S_{1}\text{-}^{3}D_{1}$ channel produces the strongest attraction, reflecting the prominent role of the tensor component of the nucleon-nucleon interaction in this coupled channel \cite{Wang_2020,Zuo_2012}. The $^{1}S_{0}$ channel provides the second-largest attractive contribution. The remaining channels include both attractive and repulsive contributions, which substantially cancel each other. Consequently, their net contribution is relatively small in the low-momentum region, although some individual channels within this group can make appreciable contributions. This behavior is consistent with the expected partial-wave hierarchy as low orbital angular momentum ($L$) channels dominate at low $k$, while higher $L$ contributions become increasingly relevant with increasing $k$, followed by an overall reduction of the interaction contributions at sufficiently high momentum. Accordingly, the magnitudes of the $^{3}S_{1}\text{-}^{3}D_{1}$ and $^{1}S_{0}$ contributions decrease with increasing momentum, while the contributions from the remaining channels become relatively more important. Thus, the pronounced hierarchy observed at low momentum becomes progressively less evident in the intermediate- and high-momentum regions.

For the subsequent analysis, we retain the $^{3}S_{1}\text{-}^{3}D_{1}$ and $^{1}S_{0}$  channels separately, while combining all remaining partial-wave contributions into a single `SUM' term,  which indicated by the dashed line in the figure.  Importantly, the relatively small SUM contribution at low momentum does not imply that the individual channels contained in it are negligible; rather, it results from cancellations among attractive and repulsive contributions. At higher momenta, these cancellations become less complete, and the collective contribution of the remaining channels can become comparable to, or even larger than, the contributions of the two individually selected channels. Table \ref{tab:other} summarizes the contributions of the $^{3}S_{1}\text{-}^{3}D_{1}$, $^{1}S_{0}$ and SUM terms to the neutron and proton single-particle potentials for several values of the asymmetry parameter and momentum. As seen from the table, increasing  $\beta$ progressively modifies the relative contributions of the different channels, reflecting the changing balance between like- and unlike-nucleon interactions. These effects are examined in more detail below through the separate neutron and proton contributions of the two dominant channels.

\begin{figure}[t!]
\centering
\includegraphics[width=\linewidth]{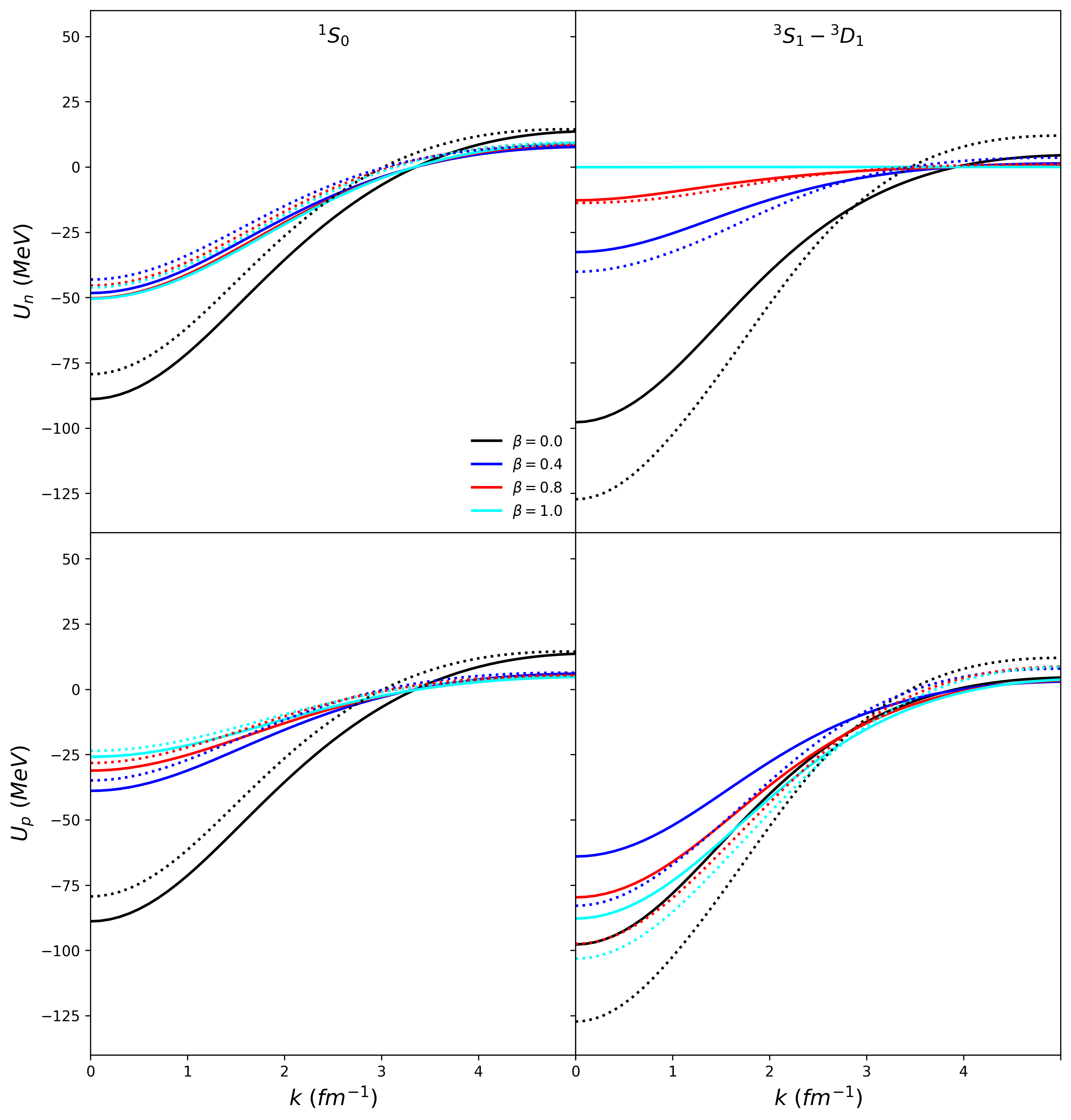}
\caption{\label{fig:5}Contributions of the remaining two channels shown in Fig. \ref{fig:4}. to the neutron (upper panel) and proton (lower panel) single-particle potential as a function of momentum for several values of the asymmetry parameter $\beta$. The results obtained with (without) the three-body force are represented by dotted (solid) lines.}
\end{figure}

Figure \ref{fig:5} shows the contributions of the two selected channels to the neutron and proton single-particle potentials as functions of momentum at $\rho=0.17~\mathrm{fm}^{-3}$ for four values of the asymmetry parameter, $\beta=0,0.4,0.8$, and $1$, with and without the three-body force. The results demonstrate that the strong dependence of the channel composition on the isospin asymmetry of the medium. The $^{3}S_{1}-{}^{3}D_{1}$ channel is a pure $T=0$ channel and therefore contributes only through neutron-proton pairs. Consequently, its contribution to the neutron SPP decreases rapidly with increasing neutron excess and vanishes in the pure-neutron-matter limit. In contrast, its contribution to the proton SPP remains finite as $\beta\rightarrow1$, since a proton can still interact with the surrounding neutrons through this channel.

 The $^{1}S_{0}$ channel belongs to the $T=1$ sector and contributes to both like-nucleon and neutron-proton interactions. Its contribution to the neutron SPP remains finite and becomes relatively more important with increasing neutron excess, reflecting the increasing abundance of neutron-neutron pairs. For the proton SPP, the contribution also remains finite in the neutron-rich limit because the proton interacts with the increasingly dominant neutron component of the medium. Thus, increasing isospin asymmetry progressively suppresses the $T=0$ contribution to the neutron SPP while enhancing the relative role of the $T=1$ contribution. This change in channel balance provides a microscopic origin for the isospin dependence of the total single-particle potentials.

The comparison with and without the TBF further shows that its effect is channel dependent. For the $^{1}S_{0}$ channel, the TBF generally reduces the attraction at low and intermediate momenta, whereas its effect on the $^{3}S_{1}-{}^{3}D_{1}$ contribution is more pronounced and can enhance the attraction at low momentum. The resulting modification of the total SPP therefore reflects the combined contribution of the different channels rather than a uniform TBF effect. As the momentum increases, the differences between the two calculations generally become smaller, showing that the impact of the TBF on the individual channel contributions is concentrated mainly in the low- and intermediate-momentum region.

\begin{figure}[b!]
\centering
\includegraphics[width=\linewidth]{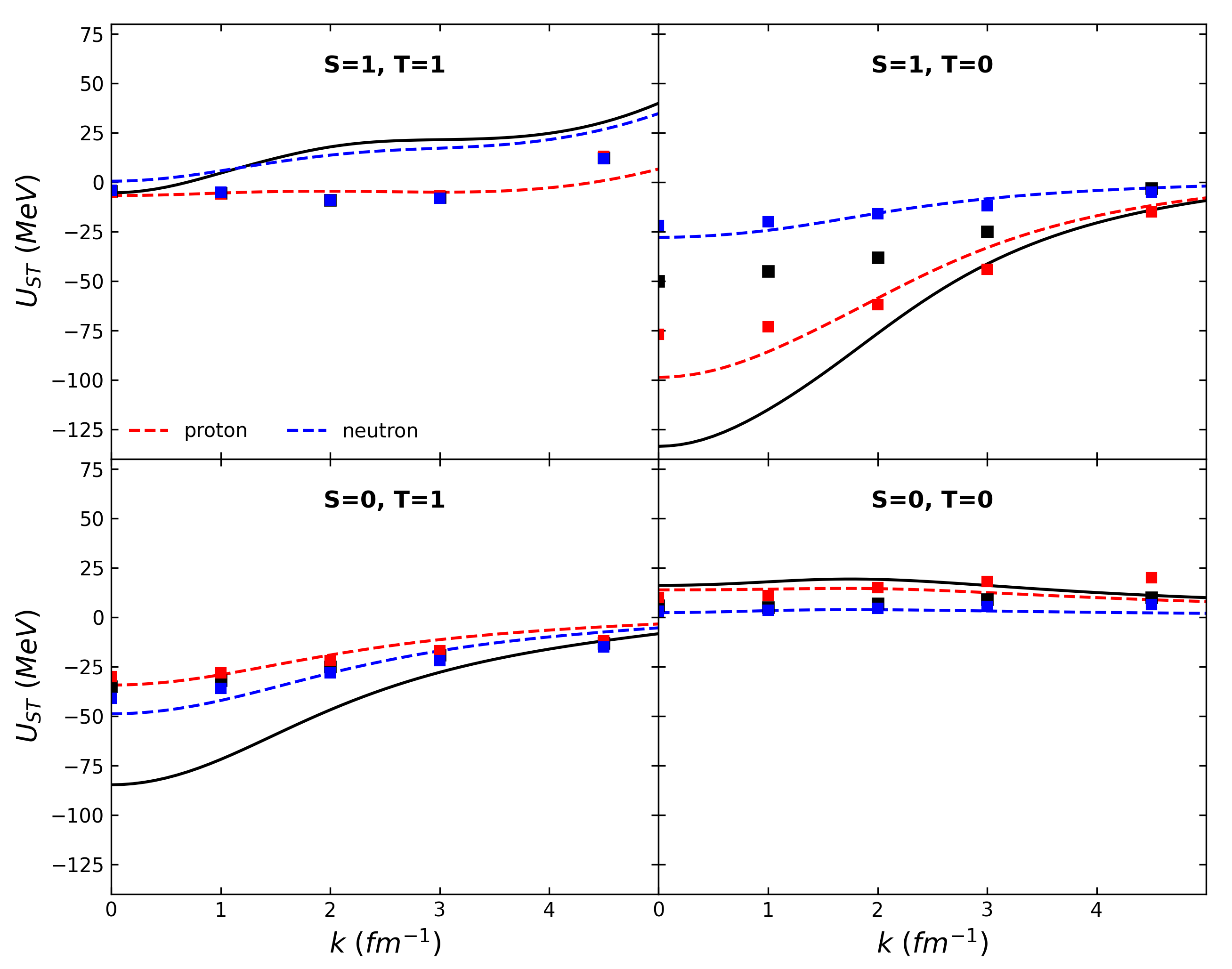}
\caption{\label{fig:6}Spin-isospin channel decomposition of the neutron (blue dashed) and proton (red dashed) single-particle potentials in asymmetric nuclear matter at $\rho = 0.17~\mathrm{fm}^{-3}$ and $\beta = 0.6$. The corresponding results for symmetric nuclear matter are indicated by the solid curves. The BHF \cite{Zuo_2012} results are also shown for comparison. ( \textcolor{red}{$\blacksquare$}  denotes the proton, \textcolor{blue}{$\blacksquare$} denotes the neutron, and \textcolor{black}{$\blacksquare$} corresponds to symmetric nuclear matter.)}
\end{figure}

We next examine the partial-wave contributions from a complementary spin-isospin perspective. Rather than considering individual $JLST$ channels, the single-particle potential is grouped into the four possible spin-isospin sectors, $(S,T)=(1,0), (0,1), (0,0)$, and $(1,1)$, where each sector contains the sum of all allowed partial-wave contributions with the corresponding values of $S$ and $T$. The resulting decomposition is shown in Fig. \ref{fig:6}. The solid curves represent symmetric nuclear matter, while the corresponding results for $\beta=0.6$ are shown for comparison.
In symmetric nuclear matter, the combined contribution of the $T=0$ and $T=1$ sectors is attractive, with the $(S,T)=(1,0)$ sector providing the largest contribution. This behavior is consistent with the strong attraction associated with the spin-triplet, isospin-singlet interaction and, in particular, with the tensor-dominated $^{3}S_{1}-{}^{3}D_{1}$ channel \cite{Zuo_2012}. The $(S,T)=(0,1)$ sector provides the next-largest contribution, while the $(S,T)=(0,0)$ and $(S,T)=(1,1)$ sectors make comparatively smaller contributions. Thus, the $(S,T)=(1,0)$ and $(S,T)=(0,1)$ sectors provide the principal contributions to the single-particle potential.

The dependence of these spin-isospin contributions on the asymmetry parameter provides further insight into the neutron-proton splitting. With increasing $\beta$, the $(S,T)=(1,0)$ contribution becomes substantially more attractive for protons, while its contribution to the neutron potential becomes much less attractive. This opposite behavior reflects the changing availability of neutron-proton pairs: a proton in neutron-rich matter continues to interact strongly with the abundant neutrons, whereas the number of neutron-proton pairs available to a neutron decreases as the proton fraction is reduced \cite{Wang_2020,Zuo_2012}. The $(S,T)=(0,1)$ sector shows a different isospin dependence, since it is associated with like-nucleon interactions. Its contribution becomes less attractive as the corresponding particle fraction decreases, with the effect being particularly evident for protons in neutron-rich matter. The $(S,T)=(0,0)$ and $(S,T)=(1,1)$ sectors remain smaller, but their different asymmetry dependence contributes to the partial cancellation among the various sectors.

Overall, the decomposition shows that the neutron-proton splitting of the SPP results from the combined effect of the spin-isospin structure of the interaction and the changing neutron and proton densities. In particular, the strong and opposite response of the $(S,T)=(1,0)$ sector for neutrons and protons demonstrates why neutron-proton correlations play a central role in the isospin dependence of the single-particle potential. The comparison with the BHF results also shows a similar overall hierarchy of the spin-isospin contributions, supporting the microscopic interpretation obtained here.

\begin{table*}[t]
\caption{\label{tab:other}The contributions of the \(^{3}S_{1}\text{--}^{3}D_{1}\), \(^{1}S_{0}\), and "SUM" channels to the proton and neutron single-particle potentials at a baryon density of $\rho = 0.17 fm^{-3}$ are presented in the table below for four asymmetry parameters, \(\beta\), and five momentum values \(k\), with the three-body force included.}
\centering
\renewcommand{\arraystretch}{1.5}

\begin{tabular}{c|c|cc|cc|cc}
\hline
\multirow{2}{*}{$\beta$} & \multirow{2}{*}{$k\,(\mathrm{fm}^{-1})$}
& \multicolumn{2}{c|}{ $^1S_0$}
& \multicolumn{2}{c|}{$^3S_1$-$^3D_1$}
& \multicolumn{2}{c}{Sum of other channels} \\
\cline{3-8}
& & $U_p(MeV)$ & $U_n(MeV)$ &$ U_p(MeV)$ &$ U_n(MeV)$ & $U_p(MeV)$ & $U_n(MeV)$ \\
\hline
\multirow{6}{*}{0}
& 0 &-79.37 &-79.36 & -127.27&-127.27 &-0.84 &-0.84 \\
& 1 &-61.48 &-61.48 & -102.44&-102.44 & -0.43& -0.43\\
& 2 & -26.42&-26.42 &-52.48 &-52.48 & -7.36&-7.36 \\
& 3 &-0.38&-0.38 &-14.06 &-14.06 & -20.11&-20.11 \\
& 4 &11.81 &11.81 & 7.82&7.82 &-19.08 &-19.08\\
& 5 &14.29 &14.39 & 12.00&12.00 &5.87 &5.87 \\ \hline
\multirow{6}{*}{0.4}
& 0 & -34.95&-43.12 &-82.92 &-40.14 & -2.16&-0.98 \\
& 1 & -27.04&-33.78 &-67.10 &-32.47 &-5.57 &1.77 \\
& 2 & -11.76&-14.94 &-35.26 &-16.23 & -14.43&2.50 \\
& 3 &-0.35 &-0.44 &-8.20 &-3.34 & -23.32&-0.94 \\
& 4 &5.09 &6.51 &4.76 &2.33 & -22.90&1.29 \\
& 5 &6.32 &8.06 &7.90 &3.56 & -9.63&16.90 \\ \hline
\multirow{6}{*}{0.8}
& 0 &-28.24 &-45.43&-97.51 &-13.77 &-4.71 &-3.61 \\
& 1 &-22.27 &-36.33 &-79.91 &-11.43 & -9.87& -2.56\\
& 2 &-10.37 &-17.04 & -43.45&-5.50 & -21.04&-2.91 \\
& 3 &-0.83 &-1.67 &-11.83 &-1.21 & -30.47&-4.18 \\
& 4 &4.06 &6.91 &4.28 &0.65 & -28.85&1.13 \\
& 5 & 5.38&8.94 &8.73&1.08 &-12.37 &20.01 \\ \hline
\multirow{6}{*}{1}
& 0 &-23.61 &-46.20 &-103.21 &0.00 & 0.92&-5.89 \\
& 1 & -19.27&-37.56 &-85.27 &0.00 & -2.94&-6.35 \\
& 2 &-9.60 &-18.33 &-47.32 &0.00 & -10.63&-7.74 \\
& 3 & -1.14&-1.76 &-14.05 &0.00 & -16.31&-8.03 \\
& 4 &3.47 &7.00 &3.42 &0.00 &-13.32 & -0.65\\
& 5 &4.87 &9.40 &8.62 &0.00 & 0.75&20.84 \\ \hline
\end{tabular}
\label{tab:results}
\end{table*}


\subsection{Symmetry potential}
\label{Symmetry potential}

Isospin symmetry allows the expansion coefficients of the single-nucleon potential to be defined as :
\begin{equation}
\begin{split}
\label{eq:21}
U_{\text{sym},i}(\rho,k) \equiv & \frac{1}{i!} \left.\frac{\partial^i U_n(\rho,\beta,k)}{\partial \beta^i}\right|_{\beta=0} \\
= & \frac{(-1)^i}{i!} \left.\frac{\partial^i U_p(\rho,\beta,k)}{\partial \beta^i}\right|_{\beta=0}.
\end{split}
\end{equation}
The single-nucleon potential can then be formally expanded in powers of the asymmetry parameter $\beta$ as \cite{PhysRevC.81.064612, PhysRevC.85.024305}
\begin{equation}
\begin{split}
U_{\tau}(\rho, \beta, k) = {} & U_{0}(\rho, k) + U_{\text{sym},1}(\rho, k)(\pm \beta) \\
& + U_{\text{sym},2}(\rho, k)(\pm \beta)^2 + \dots ,
\end{split}
\end{equation}
where $U_0(\rho,k)$ denotes the single particle potential in symmetric nuclear matter, common to neutrons and protons, and $U_{\text{sym},1}(\rho,k)$ is the first-order symmetry potential. Upon retaining only the leading-order asymmetry term, the single-particle potentials reduce to the well-known Lane approximation \cite{Lane1962},
\begin{equation}
U_{\tau}(\rho,\beta,k)\approx U_{0}(\rho,k)
+U_{\text{sym},1}(\rho,k)(\pm\beta),
\label{Ulane}
\end{equation}
where $+$ and $-$ correspond to neutrons and protons, respectively.

\begin{figure}[t!]
\centering
\includegraphics[width=\linewidth]{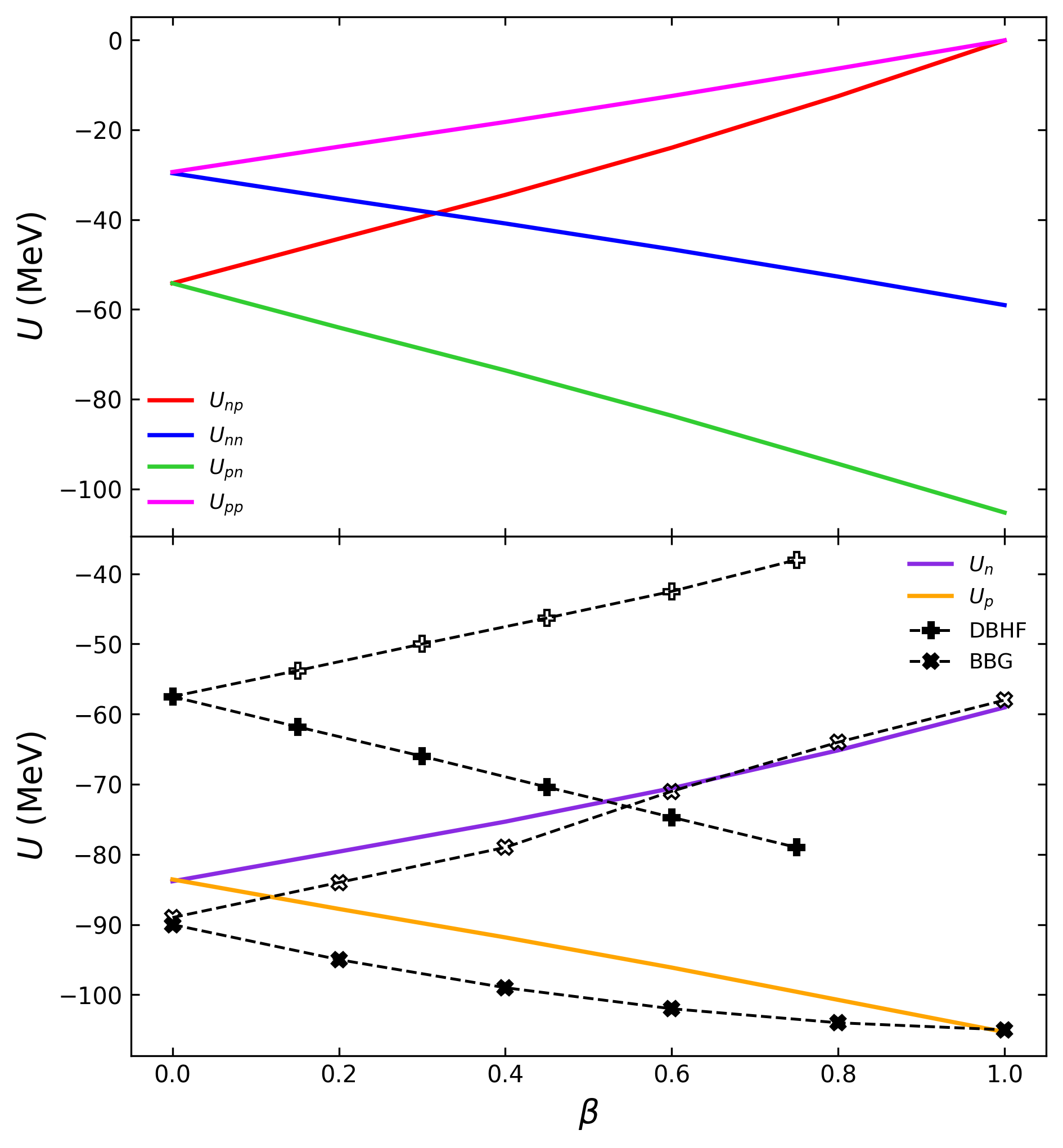}
\caption{\label{fig:7} Proton and neutron single-particle potentials as functions of the asymmetry parameter $\beta$ at $\rho=0.17~\mathrm{fm}^{-3}$ and $k=1~\mathrm{fm}^{-1}$. The $U_p$ and $U_n$ curves in the lower panel are obtained from the sums of the corresponding contributions shown in the upper panel. Results from the DBHF \cite{Sammarruca_2005} (Bonn-B) and BBG \cite{bombaci1991asymmetric} approaches are also shown for comparison. Open and filled markers correspond to neutron and proton SPPs, respectively.}
\end{figure}
Figure \ref{fig:7} displays the proton and neutron single-particle potentials at a fixed momentum $k=1~\text{fm}^{-1}$ and density $\rho=0.17~\text{fm}^{-3}$ as functions of the asymmetry parameter $\beta$ using Eq.~\ref{eq:18}. The calculated potentials exhibit an approximately linear dependence on $\beta$, consistent with the leading-order behavior expected from the Lane approximation. The results are also compared with the corresponding predictions of the BBG \cite{bombaci1991asymmetric} and DBHF \cite{Sammarruca_2005} approaches, showing consistent trends. As shown in the lower panel, the total proton and neutron single-particle potentials are obtained from the sums given in Eq. \ref{eq:18}, with their constituent contributions displayed in the upper panel. As $\beta$ increases toward pure neutron matter, the proton-related contributions, $U_{pp}$ and $U_{np}$, decrease and eventually vanish at $\beta=1$, reflecting the depletion of protons in the medium. In contrast, the neutron-related contributions, $U_{nn}$ and $U_{pn}$, increase in magnitude with increasing $\beta$ as the neutron fraction grows.

We next examine the first-order symmetry potential,
$U_{\text{sym},1}$, which characterizes the isovector component of the
single-nucleon potential. Within the first-order Lane approximation, it
can be evaluated from the difference between the neutron and proton
single-particle potentials as
\begin{equation}
U_{\text{sym},1}(\rho,k) \approx
\frac{U_n(\rho,\beta,k)-U_p(\rho,\beta,k)}{2\beta}.
\label{Usym}
\end{equation}
Since Eq.~\ref{Usym} is evaluated at finite values of $\beta$, rather
than through the derivative at $\beta=0$, the extracted quantity can in
principle contain contributions from higher-order terms in the
isospin expansion. The relatively weak dependence of the results on
$\beta$ therefore also provides an indication of the validity of the
first-order Lane approximation over the considered range of asymmetry.

\begin{figure}[b!]
\centering
\includegraphics[width=\linewidth]{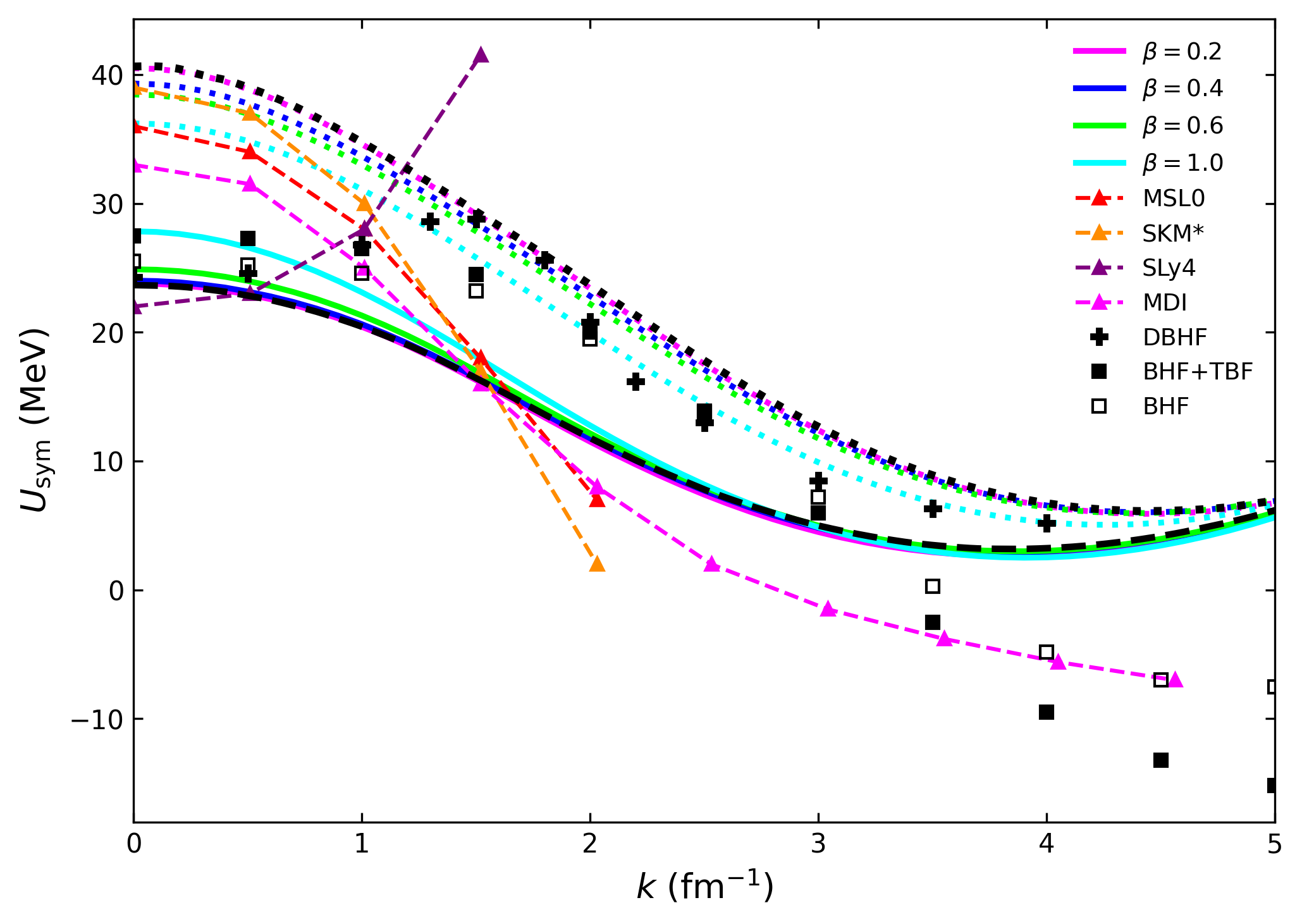}
\caption{\label{fig:8} Symmetry potential as a function of momentum at $\rho=0.17~\mathrm{fm}^{-3}$ for four values of the asymmetry parameter $\beta$. Solid and dotted lines denote calculations without and with the three-body force (TBF), respectively. Furthermore, the symmetry potential obtained from Eq.~(\ref{eq:21}) is shown in the figure, where the black dotted (dashed) line corresponds to calculations with (whthout) three-body force. Results from the DBHF \cite{Dalen_2005}, BHF \cite{PhysRevC.74.014317}, MDI, and Skyrme Hartree--Fock approaches \cite{PhysRevC.85.024305} are also shown for comparison.}
\end{figure}

Figure~\ref{fig:8} presents the symmetry potential as a function of
momentum at the baryon density $\rho=0.17~\mathrm{fm}^{-3}$ for several
values of the asymmetry parameter $\beta$. The symmetry potential
decreases systematically with increasing momentum, indicating that the
isovector part of the single-nucleon potential becomes progressively
weaker at higher momenta. The inclusion of the TBF does not alter this
qualitative momentum dependence but significantly enhances the
symmetry potential at low momentum. At $k=0$, the inclusion of the TBF
increases $U_{\text{sym},1}$ by about $16$~MeV. This difference
gradually decreases with increasing momentum and becomes nearly
negligible at $k\sim5~\mathrm{fm}^{-1}$. Thus, the TBF mainly modifies
the low-momentum part of the isovector potential, while its influence
becomes progressively weaker at high momentum.

For comparison, results from the DBHF \cite{Dalen_2005}, BHF
\cite{PhysRevC.74.014317}, MDI, and Skyrme Hartree--Fock
\cite{PhysRevC.85.024305} approaches are also shown. The decreasing
momentum dependence obtained in the present calculation is broadly
consistent with the BHF, DBHF, and Skyrme Hartree--Fock results, as
well as with the MDI interaction, while the SLy4
parametrization exhibits an opposite trend
\cite{PhysRevC.85.024305}.

To assess the accuracy of the Lane approximation, the exact first-order symmetry potential obtained directly from Eq. (\ref{eq:21}) is also shown by the black dashed and dotted lines correspond to calculations without and with the three-body force, respectively. Its close agreement with the results extracted from Eq. (24) for the different finite values of $\beta$ demonstrates that the Lane approximation provides a reliable representation of the first-order symmetry potential over the momentum and asymmetry ranges considered here. The relatively weak dependence of the
symmetry potential on $\beta$, particularly at higher momenta, further
supports the adequacy of the first-order approximation in this regime,
whereas the deviations at low momentum and larger asymmetry indicate
that finite-asymmetry effects become more noticeable there.

\begin{figure}[t]
\centering
\includegraphics[width=\linewidth]{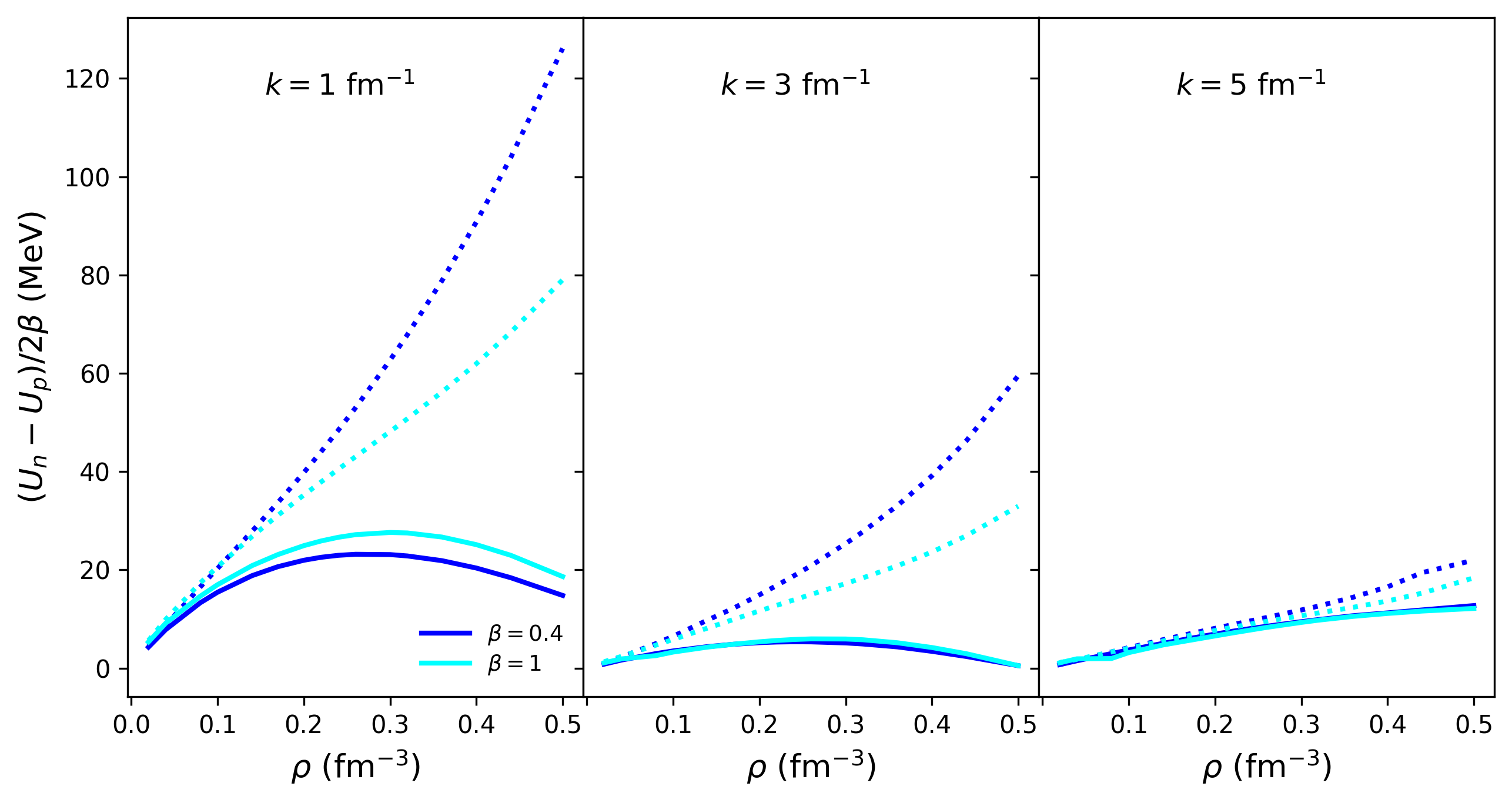}
\caption{\label{fig:9}The symmetry potential as a function of density is plotted at three fixed momenta and for two asymmetry parameters. The results obtained with and without the three-body force are shown by dotted and solid lines, respectively.}
\end{figure}

Figure~\ref{fig:9} shows the symmetry potential as a function of
density for three fixed momenta and two asymmetry parameters,
$\beta=0.4$ and $1$. The calculations with and without the
three-body force are represented by dotted and solid lines,
respectively. The results further demonstrate the pronounced
state dependence of the symmetry potential: its density dependence
changes substantially with momentum. At low momentum, the symmetry
potential increases with density , reaches a maximum
around $\rho\simeq0.3~\mathrm{fm}^{-3}$, and subsequently decreases
at higher densities. This nonmonotonic behavior becomes progressively
weaker with increasing momentum, and at high momentum the density
dependence becomes approximately monotonic. Thus, the density and
momentum dependences of the symmetry potential are strongly coupled
rather than independent.

The inclusion of the TBF enhances the symmetry potential, with its
effect being particularly pronounced at low momentum and increasing
density. As the momentum increases, the difference between the
calculations with and without the TBF becomes progressively smaller.
This behavior is consistent with that observed in Fig.~\ref{fig:8},
where the TBF contribution is strongest at low momentum and approaches
negligible values at high momentum. The dependence of the results on
$\beta$ remains relatively weak over a substantial part of the
considered parameter space, although larger deviations are visible at
low momentum and high asymmetry, again reflecting the increasing
importance of finite-asymmetry contributions beyond the leading-order
Lane term.

An important feature of the present approach is that the symmetry
potential is not introduced through a phenomenological parametrization.
Instead, it is obtained directly from the microscopic LOCV
single-particle potentials and therefore retains the state dependence
generated by the underlying realistic two- and three-body interactions.
Its momentum, density, and asymmetry dependence consequently emerge
from the microscopic calculation itself, providing a state-dependent
single-particle potential that can serve as input for applications where
phenomenological or mean-field-based parametrizations are commonly
employed.

Empirical analyses of nucleon--nucleus scattering at energies below
$100~\mathrm{MeV}$ indicate that the Lane potential decreases
approximately linearly with increasing kinetic energy, $E_{\mathrm{kin}}$,
and can be parametrized as \cite{LANE1962676,Li2004}
\begin{equation}
U_{\mathrm{Lane}}=a-bE_{\mathrm{kin}},
\label{Ulane2}
\end{equation}
\begin{figure}[b]
\centering
\includegraphics[width=\linewidth]{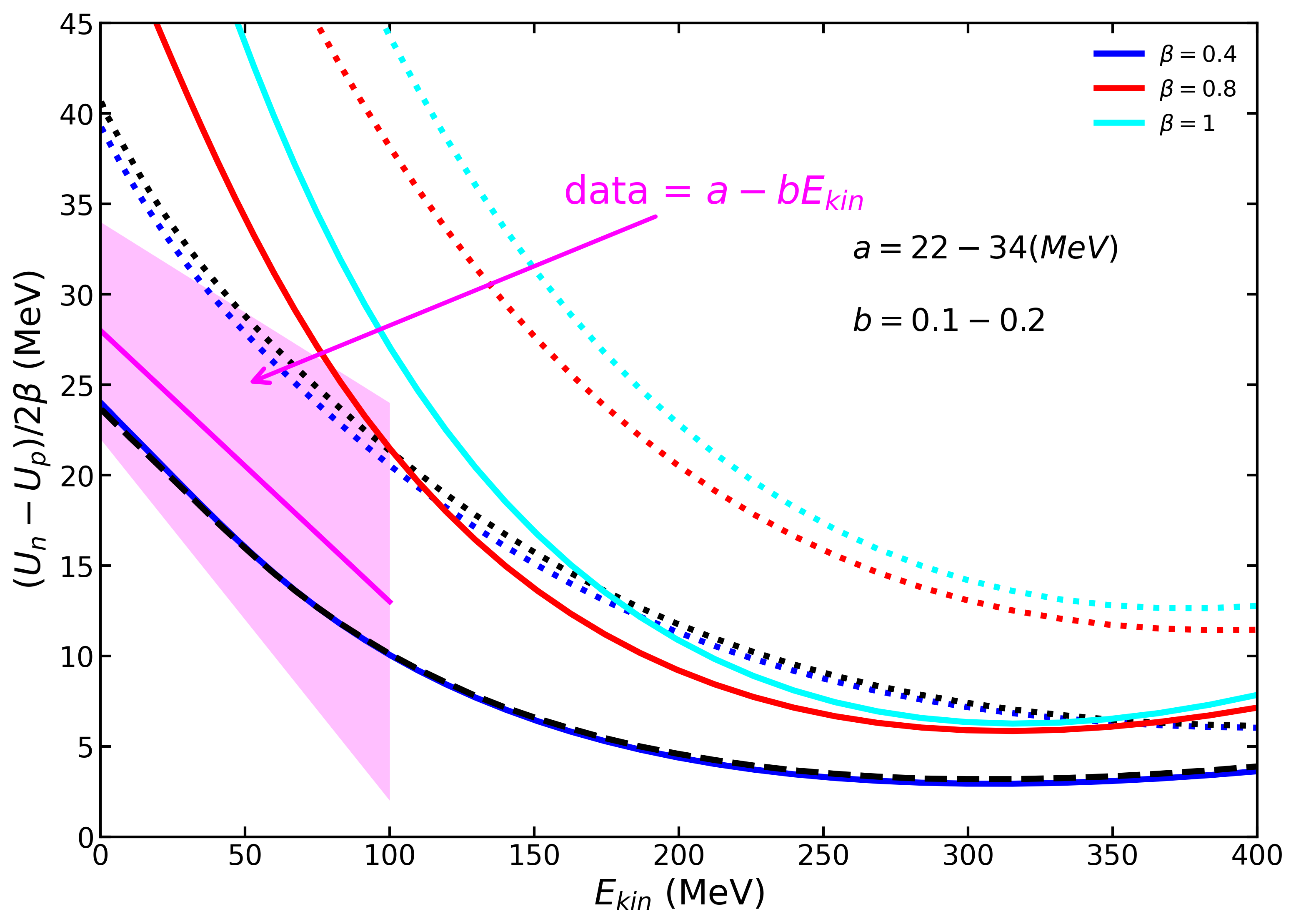}
\caption{\label{fig:10}Symmetry potential as a function of kinetic energy at
$\rho=0.17~\mathrm{fm}^{-3}$ for three values of the asymmetry parameter
$\beta$. Solid and dotted curves denote calculations without and with the
three-body force (TBF), respectively. Furthermore, the results obtained from Eq.~(\ref{eq:21}) are also presented, where the black dotted (dashed) line corresponds to calculations with (without) the three-body force. The shaded region represents the
empirically constrained range of the Lane potential
$U_{\mathrm{Lane}}=a-bE_{\mathrm{kin}}$, with
$a=22$--$34~\mathrm{MeV}$ and $b=0.1$--$0.2$ \cite{Lane1962,Li2004}.}
\end{figure}
where $a\approx22-34~\mathrm{MeV}$ and $b\approx0.1-0.2$.
Figure \ref{fig:10} compares the calculated symmetry potential at
$\rho=0.17~\mathrm{fm}^{-3}$ for three values of the asymmetry parameter
$\beta$, {as well as calculated from Eq.~(\ref{eq:21})} with the empirically constrained range of the Lane potential.
The shaded region represents the full parameter space spanned by the
allowed ranges of $a$ and $b$. In all cases, the calculated symmetry
potential decreases with increasing $E_{\mathrm{kin}}$, reproducing the
empirical decreasing trend. Among the considered asymmetries, the results
for $\beta=0.4$ and {direct calculation from eq.~(\ref{eq:21})} lie within the empirical constraint, whereas the other
two asymmetries remain outside the allowed region. The inclusion of the
TBF makes the symmetry potential more repulsive, particularly at low
kinetic energies, consistent with the behavior observed in
Fig.~\ref{fig:8}. For $\beta=0.4$ {and also $U_{sym,1}$ directly obtained from our approache}, the TBF-included result enters the
empirically constrained region for $E_{\mathrm{kin}}\gtrsim40~\mathrm{MeV}$,
while its effect becomes progressively smaller at higher energies.

\subsection{Effective mass}
\label{Effective mass}

The nucleon effective mass is an important quantity characterizing the momentum dependence of the single-particle potential. In asymmetric nuclear matter, the neutron-proton effective-mass splitting is closely
related to the momentum dependence of the isovector single-particle potential and, consequently, to the energy dependence of the symmetry potential \cite{Li_2015}. The effective mass is determined from the
momentum derivative of the single-particle energy as
\begin{equation}
\frac{m^{*}}{m}
=
\frac{k}{m}
\left[\frac{de(k)}{dk}\right]^{-1}.
\end{equation}
For each nucleon species $\tau$, this relation can be written explicitly
in terms of the corresponding single-particle potential as
\begin{equation}
\frac{m_\tau^{*}}{m}
=
\left[
1+\frac{m}{k}
\frac{\partial U_\tau(\rho,\beta,k)}{\partial k}
\right]^{-1}.
\end{equation}

\begin{figure}[t!]
\centering
\includegraphics[width=\linewidth]{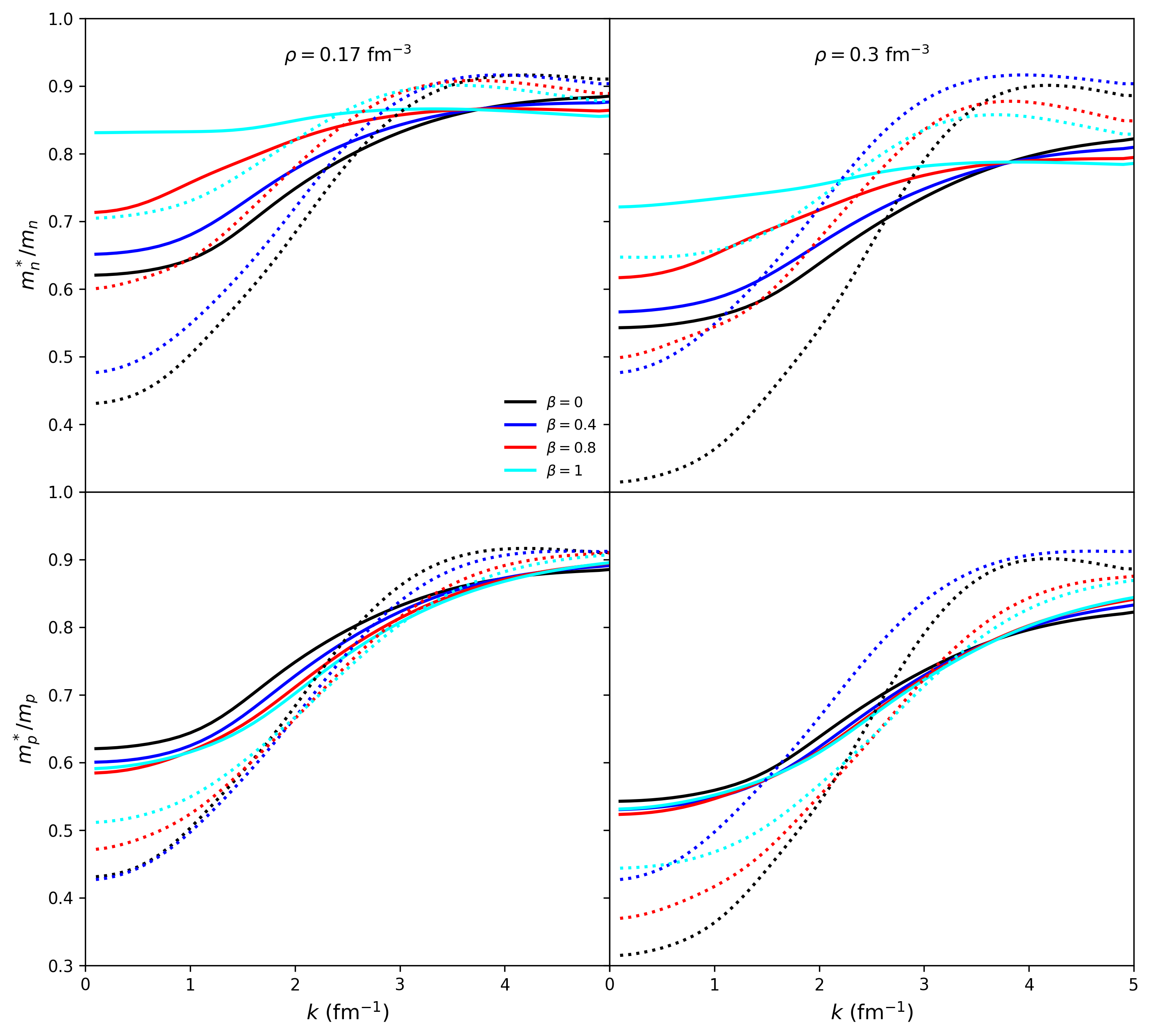}
\caption{\label{fig:11}Neutron (upper panels) and proton (lower panels) effective masses as functions of momentum at the baryon densities $\rho=0.17$ and 
$0.30~\mathrm{fm}^{-3}$ for four values of the asymmetry parameter ($\beta$). Dotted and solid curves correspond to calculations with and without the three-body force, respectively.}
\end{figure}

Figure \ref{fig:11} shows the neutron and proton effective masses as functions of momentum at the baryon densities $\rho=0.17$ and $0.30~\mathrm{fm}^{-3}$ for several values of the asymmetry parameter
$\beta$. The results obtained with and without the three-body force are represented by do0tted and solid lines, respectively. The effective masses exhibit a clear dependence on both momentum and isospin asymmetry. As the isospin asymmetry increases, the neutron effective mass becomes larger than the proton effective mass, yielding a neutron-proton effective-mass splitting consistent with nonrelativistic microscopic calculations \cite{bombaci1991asymmetric}. This splitting originates from the different momentum dependences of the neutron and proton single-particle potentials in asymmetric matter. The momentum dependence of the splitting is also evident. At low and intermediate momenta, the neutron and proton effective masses become increasingly separated as the asymmetry increases, whereas their difference becomes smaller at larger momenta. Thus, the isospin splitting of the effective mass is mainly associated with the momentum dependence of the mean field in the low- and intermediate-momentum region.

The inclusion of the three-body force produces an additional and strongly density-dependent modification of the effective masses. In the low- and intermediate-momentum region, the TBF generally reduces the effective masses of both nucleon species, indicating a stronger momentum dependence of the corresponding single-particle potentials. This effect becomes considerably more pronounced at $\rho=0.3 fm^{-3}$, reflecting the increasing importance of three-body correlations with density. At larger momenta, however, the difference between the results with and without the TBF becomes smaller, and the corresponding effective masses tend to approach each other. This behavior is also consistent with the momentum dependence of the symmetry potential shown in Fig.~\ref{fig:8}. The TBF produces its largest modification of the symmetry potential at low momentum, while the results with and without the TBF gradually converge as the momentum increases. The effective-mass results therefore provide a complementary manifestation of the momentum- and density-dependent effects introduced by the three-body force.

\begin{figure}[H]
\centering
\includegraphics[width=\linewidth]{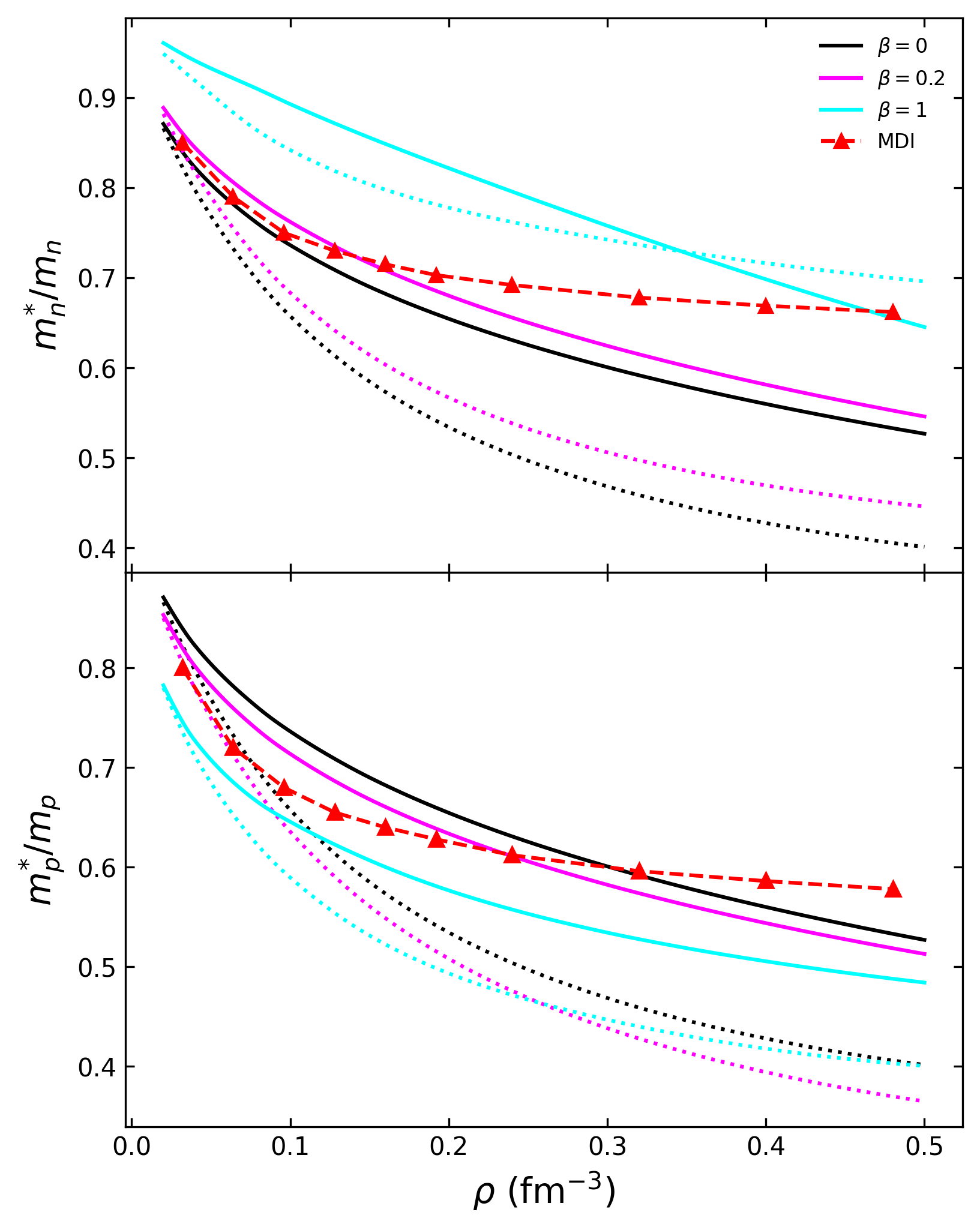}
\caption{\label{fig:12}
Density dependence of the neutron (upper panel) and proton (lower panel)
effective masses, evaluated at their respective Fermi momenta $k_{F_n}$
and $k_{F_p}$, for three values of the asymmetry parameter $\beta$.
Solid and dotted curves correspond to calculations without and with the
three-body force, respectively. Results obtained with the MDI interaction
\cite{LiChen2005} are also shown for comparison.}
\end{figure}

Figure \ref{fig:12} shows the density dependence of the neutron and proton effective masses, evaluated at their respective Fermi momenta $k_{F_n}$ and $k_{F_p}$, for three values of the asymmetry parameter $\beta=0$, $0.2$, and $1$. The results obtained with and without the
three-body force are represented by dotted and solid lines, respectively. For both nucleon species, the effective masses decrease with increasing density, reflecting the increasing momentum dependence of the single-particle potentials in denser matter. The inclusion of the TBF further modifies the density dependence of the effective masses. Its effect is relatively weak at low density but becomes increasingly important as the density increases, leading to a substantial reduction of the effective masses at suprasaturation densities. This behavior is consistent with the results of Fig. \ref{fig:11} and reflects the growing role of three-body correlations in determining the momentum dependence of the nucleon mean field at high density. We can see the isospin asymmetry dependence on effective masses in the figure too. At a given density, the effective masses of neutrons and protons respond differently to increasing $\beta$, leading to the neutron-proton effective-mass splitting discussed above. For comparison, results obtained using the momentum-dependent isospin-dependent (MDI) interaction are also shown. The present results exhibit an overall behavior consistent with the MDI predictions \cite{LiChen2005,Li_2015}, particularly regarding the decrease of the effective masses with density and their dependence on the isospin asymmetry, although quantitative differences are observed for some densities and asymmetries.

\begin{figure}[t]
\centering
\includegraphics[width=\linewidth]{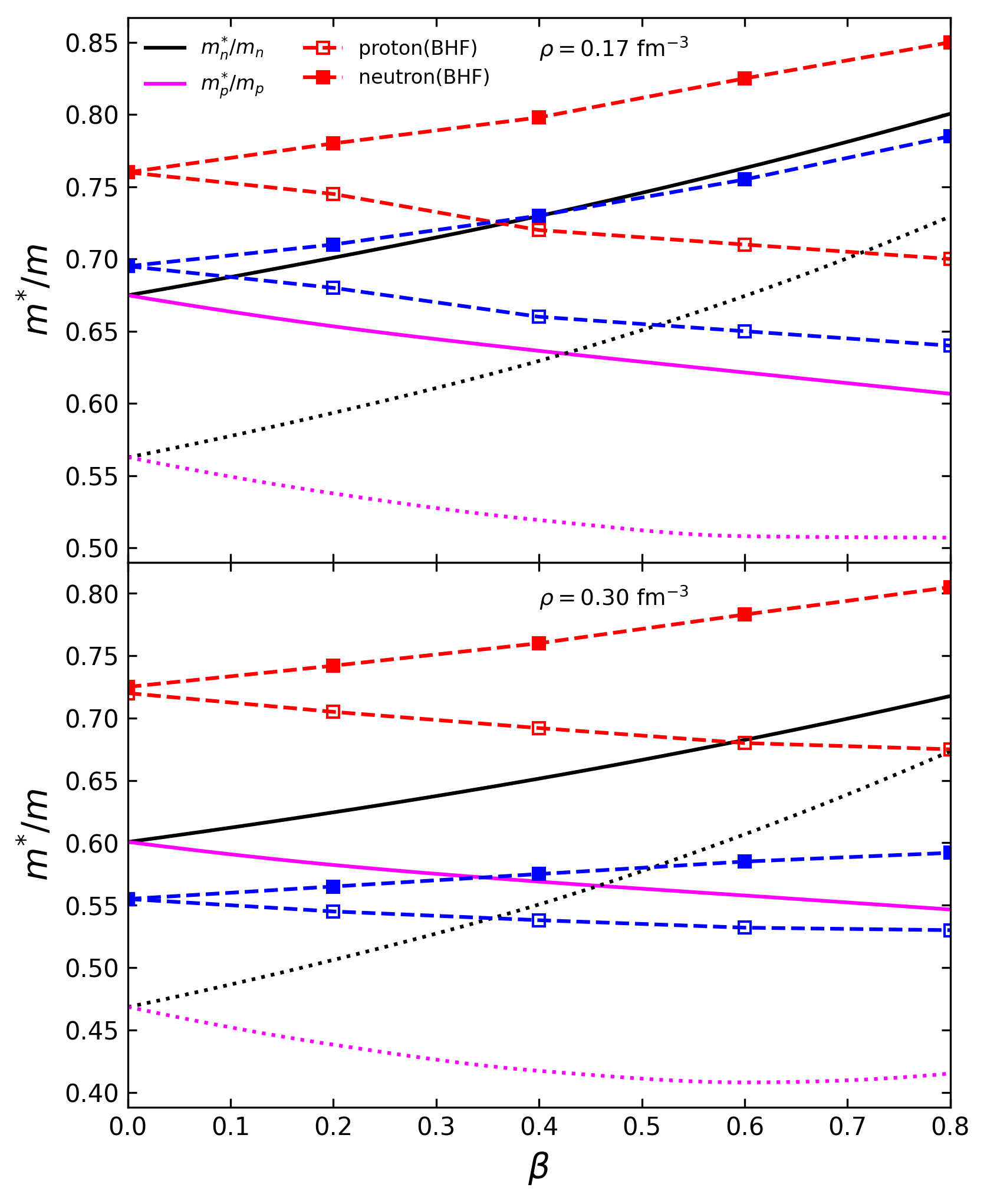}
\caption{\label{fig:13}
Neutron and proton effective masses as functions of the asymmetry
parameter $\beta$ at $\rho=0.17$ and $0.3~\mathrm{fm}^{-3}$, evaluated at
their respective Fermi momenta. Solid and dotted curves correspond to
calculations without and with the three-body force, respectively. BHF
results \cite{PhysRevC.74.014317} are also shown for comparison; blue
(red) squares denote calculations with (without) the three-body force.}
\end{figure}

The dependence of the nucleon effective masses on the asymmetry parameter
$\beta$ is shown in Fig. \ref{fig:13} for the baryon densities
$\rho=0.17$ and $0.3~\mathrm{fm}^{-3}$, evaluated at their respective
Fermi momenta. As the system becomes increasingly asymmetric, a
splitting develops between the neutron and proton effective masses, with
$m_n^*>m_p^*$ and the magnitude of the splitting generally  increases with
$\beta$. This behavior is consistent with results obtained using other
many-body approaches
\cite{PhysRevC.74.014317,Li_2015,LiChen2005,bombaci1991asymmetric,Sammarruca_2005}.
The inclusion of the three-body force reduces the effective masses of both
nucleon species. However, its effect is not identical for neutrons and
protons. The reduction is generally stronger for the proton effective mass,
leading to an enhancement of the neutron-proton effective-mass splitting.
This effect is more pronounced at the higher density, where the TBF produces
a larger separation between the neutron and proton effective masses. The different response of the neutron and proton effective masses to the three-body force can be understood from the momentum dependence of the
corresponding single-particle potentials. In asymmetric matter, neutrons and
protons experience different surrounding media and sample the spin-isospin
components of the interaction with different weights. The density-dependent
contribution generated by the three-body force therefore modifies
$\partial U_n/\partial k$ and $\partial U_p/\partial k$ differently. This
leads to a species-dependent reduction of the effective masses and can modify
the neutron-proton effective-mass splitting. The effect becomes more important
at higher density, where three-body correlations play a larger role. For
comparison, the BHF results obtained with and without the three-body force
are also shown \cite{PhysRevC.74.014317}. The overall behavior of the
present results is consistent with the BHF predictions, although quantitative
differences remain due to the different treatments of correlations and
three-body effects.
\section{conclusion}
\label{sec:conclusion}

In this work, we have investigated the single-particle potential (SPP) of
neutrons and protons in asymmetric nuclear matter within the LOCV
framework. Realistic two-body interactions and three-body forces were
included. The SPP was studied as a function of density, momentum, and
isospin asymmetry. We also analyzed its contributions from different
spin-isospin channels and examined the resulting symmetry potential and
nucleon effective masses.

The neutron and proton SPPs show a clear splitting with increasing
asymmetry. The channel decomposition shows that the $^3S_1$--$^3D_1$ and
$^1S_0$ channels provide the dominant attractive contributions, while
other channels give smaller contributions or partially cancel each
other. The symmetry potential decreases with increasing momentum. The
three-body force makes it more repulsive, particularly at low momentum,
while its effect becomes small at high momentum. Its density dependence
is also found to depend strongly on momentum, demonstrating the
state-dependent nature of the microscopic SPP.

The neutron-proton effective-mass splitting increases with isospin
asymmetry, with $m_n^*>m_p^*$ in the present calculation. The inclusion
of the three-body force reduces both effective masses but does not
substantially change their splitting. The calculated symmetry potential
also reproduces the decreasing trend of the empirical Lane potential
with kinetic energy. In particular, the result for $\beta=0.4$ lies
within the empirically constrained region, providing an additional
consistency check on the present microscopic calculation.

An important feature of the present approach is that the SPP and its
isovector component are obtained directly from the microscopic LOCV
calculation rather than introduced through a phenomenological
parametrization. Their momentum, density, and asymmetry dependence
therefore emerge from the underlying two- and three-body interactions.
The resulting state-dependent SPP can provide microscopic input for
studies in which parametrized single-particle potentials are commonly
used.

Several extensions of the present work are of interest. An extension to
finite temperature would allow the temperature dependence of the SPP,
symmetry potential, and effective masses to be studied within the same
microscopic framework. The present results can also be extended to
finite nuclei, providing a possible microscopic basis for investigating
properties such as neutron-skin thicknesses.

\bibliography{reff}

@PREAMBLE{
 "\providecommand{\noopsort}[1]{}" 
 # "\providecommand{\singleletter}[1]{#1}%" 
}

@article{Russotto:2023ari,
    author = "Russotto, P. and Cozma, M. D. and De Filippo, E. and F{\`e}vre, A. Le and Leifels, Y. and {\L}ukasik, J.",
    title = "{Studies of the equation-of-state of nuclear matter by heavy-ion collisions at intermediate energy in the multi-messenger era: A review focused on GSI results}",
    eprint = "2302.01453",
    archivePrefix = "arXiv",
    primaryClass = "nucl-ex",
    doi = "10.1007/s40766-023-00039-4",
    journal = "Riv. Nuovo Cim.",
    volume = "46",
    number = "1",
    pages = "1--70",
    year = "2023"
}

@inbook{HAGINO_2013,
   title={Exotic Nuclei Far from the Stability Line},
   ISBN={9789814425810},
   url={http://dx.doi.org/10.1142/9789814425810_0009},
   DOI={10.1142/9789814425810_0009},
   booktitle={100 Years of Subatomic Physics},
   publisher={WORLD SCIENTIFIC},
   author={Hagino, K. and Tanihata, I. and Sagawa, H.},
   year={2013},
 pages={231–272} }

@article{Li_1998,
   title={Isospin Physics in Heavy-Ion Collisions at Intermediate Energies},
   volume={07},
   ISSN={1793-6608},
   url={http://dx.doi.org/10.1142/S0218301398000087},
   DOI={10.1142/s0218301398000087},
   number={02},
   journal={International Journal of Modern Physics E},
   publisher={World Scientific Pub Co Pte Lt},
   author={Li, Bao-An and Ko, Che Ming and Bauer, Wolfgang},
   year={1998},
   month=Apr, pages={147–229} }

@article{DANIELEWICZ2003233,
title = {Surface symmetry energy},
journal = {Nuclear Physics A},
volume = {727},
number = {3},
pages = {233-268},
year = {2003},
issn = {0375-9474},
doi = {https://doi.org/10.1016/j.nuclphysa.2003.08.001},
url = {https://www.sciencedirect.com/science/article/pii/S0375947403016622},
author = {Paweł Danielewicz}
}

@article{LATTIMER2007109,
title = {Neutron star observations: Prognosis for equation of state constraints},
journal = {Physics Reports},
volume = {442},
number = {1},
pages = {109-165},
year = {2007},
note = {The Hans Bethe Centennial Volume 1906-2006},
issn = {0370-1573},
doi = {https://doi.org/10.1016/j.physrep.2007.02.003},
url = {https://www.sciencedirect.com/science/article/pii/S0370157307000452},
author = {James M. Lattimer and Madappa Prakash}
}

@article{STEINER2005325,
title = {Isospin asymmetry in nuclei and neutron stars},
journal = {Physics Reports},
volume = {411},
number = {6},
pages = {325-375},
year = {2005},
issn = {0370-1573},
doi = {https://doi.org/10.1016/j.physrep.2005.02.004},
url = {https://www.sciencedirect.com/science/article/pii/S0370157305001043},
author = {A.W. Steiner and M. Prakash and J.M. Lattimer and P.J. Ellis}
}

@article{BARAN2005335,
title = {Reaction dynamics with exotic nuclei},
journal = {Physics Reports},
volume = {410},
number = {5},
pages = {335-466},
year = {2005},
issn = {0370-1573},
doi = {https://doi.org/10.1016/j.physrep.2004.12.004},
url = {https://www.sciencedirect.com/science/article/pii/S0370157305000025},
author = {V. Baran and M. Colonna and V. Greco and M. {Di Toro}}
}

@article{LI2008113,
title = {Recent progress and new challenges in isospin physics with heavy-ion reactions},
journal = {Physics Reports},
volume = {464},
number = {4},
pages = {113-281},
year = {2008},
issn = {0370-1573},
doi = {https://doi.org/10.1016/j.physrep.2008.04.005},
url = {https://www.sciencedirect.com/science/article/pii/S0370157308001269},
author = {Bao-An Li and Lie-Wen Chen and Che Ming Ko}
}

@article{PhysRevC.85.024305,
  title = {Single-nucleon potential decomposition of the nuclear symmetry energy},
  author = {Chen, Rong and Cai, Bao-Jun and Chen, Lie-Wen and Li, Bao-An and Li, Xiao-Hua and Xu, Chang},
  journal = {Phys. Rev. C},
  volume = {85},
  issue = {2},
  pages = {024305},
  numpages = {15},
  year = {2012},
  month = {Feb},
  publisher = {American Physical Society},
  doi = {10.1103/PhysRevC.85.024305},
  url = {https://link.aps.org/doi/10.1103/PhysRevC.85.024305}
}

@article{BERTSCH1988189,
title = {A guide to microscopic models for intermediate energy heavy ion collisions},
journal = {Physics Reports},
volume = {160},
number = {4},
pages = {189-233},
year = {1988},
issn = {0370-1573},
doi = {https://doi.org/10.1016/0370-1573(88)90170-6},
url = {https://www.sciencedirect.com/science/article/pii/0370157388901706},
author = {G.F. Bertsch and S. {Das Gupta}}
}

@article{AICHELIN1991233,
title = {“Quantum” molecular dynamics—a dynamical microscopic n-body approach to investigate fragment formation and the nuclear equation of state in heavy ion collisions},
journal = {Physics Reports},
volume = {202},
number = {5},
pages = {233-360},
year = {1991},
issn = {0370-1573},
doi = {https://doi.org/10.1016/0370-1573(91)90094-3},
url = {https://www.sciencedirect.com/science/article/pii/0370157391900943},
author = {Jörg Aichelin}
}

@article{PhysRevLett.58.1926,
  title = {Importance of Momentum-Dependent Interactions for the Extraction of the Nuclear Equation of State from High-Energy Heavy-Ion Collisions},
  author = {Aichelin, J. and Rosenhauer, A. and Peilert, G. and Stoecker, H. and Greiner, W.},
  journal = {Phys. Rev. Lett.},
  volume = {58},
  issue = {19},
  pages = {1926--1929},
  numpages = {0},
  year = {1987},
  month = {May},
  publisher = {American Physical Society},
  doi = {10.1103/PhysRevLett.58.1926},
  url = {https://link.aps.org/doi/10.1103/PhysRevLett.58.1926}
}

@article{PhysRevC.40.R491,
  title = {Momentum dependence of the nuclear mean field},
  author = {Baldo, M. and Bombaci, I. and Giansiracusa, G. and Lombardo, U.},
  journal = {Phys. Rev. C},
  volume = {40},
  issue = {2},
  pages = {R491(R)--R494(R)},
  numpages = {0},
  year = {1989},
  month = {Aug},
  publisher = {American Physical Society},
  doi = {10.1103/PhysRevC.40.R491},
  url = {https://link.aps.org/doi/10.1103/PhysRevC.40.R491}
}

@article{INSOLIA199412,
title = {Nuclear dynamics for heavy ion collisions with a momentum dependent potential},
journal = {Physics Letters B},
volume = {334},
number = {1},
pages = {12-17},
year = {1994},
issn = {0370-2693},
doi = {https://doi.org/10.1016/0370-2693(94)90584-3},
url = {https://www.sciencedirect.com/science/article/pii/0370269394905843},
author = {A. Insolia and U. Lombardo and N.G. Sandulescu and A. Bonasera}
}

@article{ZUO19981,
title = {Single-particle properties in neutron matter from extended Brueckner theory},
journal = {Physics Letters B},
volume = {421},
number = {1},
pages = {1-7},
year = {1998},
issn = {0370-2693},
doi = {https://doi.org/10.1016/S0370-2693(97)01600-6},
url = {https://www.sciencedirect.com/science/article/pii/S0370269397016006},
author = {W Zuo and G Giansiracusa and U Lombardo and N Sandulescu and H.-J Schulze}
}

@article{PhysRevC.59.2927,
  title = {Superfluidity and single-particle energies in nuclear matter},
  author = {Lombardo, U. and Schulze, H.-J. and Zuo, W.},
  journal = {Phys. Rev. C},
  volume = {59},
  issue = {5},
  pages = {2927--2930},
  numpages = {0},
  year = {1999},
  month = {May},
  publisher = {American Physical Society},
  doi = {10.1103/PhysRevC.59.2927},
  url = {https://link.aps.org/doi/10.1103/PhysRevC.59.2927}
}

@article{ZUO2010574c,
title = {EOS and Single Particle Properties of Asymmetric Nuclear Matter},
journal = {Nuclear Physics A},
volume = {834},
number = {1},
pages = {574c-576c},
year = {2010},
note = {The 10th International Conference on Nucleus-Nucleus Collisions (NN2009)},
issn = {0375-9474},
doi = {https://doi.org/10.1016/j.nuclphysa.2010.01.095},
url = {https://www.sciencedirect.com/science/article/pii/S0375947410000965},
author = {Wei Zuo}
}

@article{PhysRevC.73.035208,
  title = {Temperature dependence of single-particle properties in nuclear matter},
  author = {Zuo, W. and Li, Z. H. and Lombardo, U. and Lu, G. C. and Schulze, H.-J.},
  journal = {Phys. Rev. C},
  volume = {73},
  issue = {3},
  pages = {035208},
  numpages = {8},
  year = {2006},
  month = {Mar},
  publisher = {American Physical Society},
  doi = {10.1103/PhysRevC.73.035208},
  url = {https://link.aps.org/doi/10.1103/PhysRevC.73.035208}
}

@article{PhysRevC.72.034005,
  title = {Nonlocality in the nucleon-nucleon interaction and nuclear matter saturation},
  author = {Baldo, M. and Maieron, C.},
  journal = {Phys. Rev. C},
  volume = {72},
  issue = {3},
  pages = {034005},
  numpages = {8},
  year = {2005},
  month = {Sep},
  publisher = {American Physical Society},
  doi = {10.1103/PhysRevC.72.034005},
  url = {https://link.aps.org/doi/10.1103/PhysRevC.72.034005}
}

@article{Sammarruca_2005,
   title={Predicting the single-proton and single-neutron potentials in asymmetric nuclear matter},
   volume={71},
   ISSN={1089-490X},
   url={http://dx.doi.org/10.1103/PhysRevC.71.064306},
   DOI={10.1103/physrevc.71.064306},
   number={6},
   journal={Physical Review C},
   publisher={American Physical Society (APS)},
   author={Sammarruca, F. and Barredo, W. and Krastev, P.},
   year={2005},
 }

@article{PhysRevLett.56.1237,
  title = {Equation of State of Nuclear Matter in the Relativistic Dirac-Brueckner Approach},
  author = {ter Haar, Bernard and Malfliet, Rudi},
  journal = {Phys. Rev. Lett.},
  volume = {56},
  issue = {12},
  pages = {1237--1240},
  numpages = {0},
  year = {1986},
  month = {Mar},
  publisher = {American Physical Society},
  doi = {10.1103/PhysRevLett.56.1237},
  url = {https://link.aps.org/doi/10.1103/PhysRevLett.56.1237}
}

@article{PhysRevC.48.2707,
  title = {Momentum-dependent mean field based upon the Dirac-Brueckner approach for nuclear matter},
  author = {Li, G. Q. and Machleidt, R.},
  journal = {Phys. Rev. C},
  volume = {48},
  issue = {6},
  pages = {2707--2713},
  numpages = {0},
  year = {1993},
  month = {Dec},
  publisher = {American Physical Society},
  doi = {10.1103/PhysRevC.48.2707},
  url = {https://link.aps.org/doi/10.1103/PhysRevC.48.2707}
}

@article{LEE1997235,
title = {Momentum dependence of single particle potential in Dirac Brueckner approach},
journal = {Physics Letters B},
volume = {412},
number = {3},
pages = {235-239},
year = {1997},
issn = {0370-2693},
doi = {https://doi.org/10.1016/S0370-2693(97)01058-7},
url = {https://www.sciencedirect.com/science/article/pii/S0370269397010587},
author = {C.-H. Lee and T.T.S. Kuo and G.Q. Li and G.E. Brown}
}

@article{PhysRevC.72.065803,
  title = {Momentum, density, and isospin dependence of symmetric and asymmetric nuclear matter properties},
  author = {Dalen, E. N. E. van and Fuchs, C. and Faessler, Amand},
  journal = {Phys. Rev. C},
  volume = {72},
  issue = {6},
  pages = {065803},
  numpages = {9},
  year = {2005},
  month = {Dec},
  publisher = {American Physical Society},
  doi = {10.1103/PhysRevC.72.065803},
  url = {https://link.aps.org/doi/10.1103/PhysRevC.72.065803}
}

@article{PhysRevC.76.054316,
  title = {Isospin-dependent properties of asymmetric nuclear matter in relativistic mean field models},
  author = {Chen, Lie-Wen and Ko, Che Ming and Li, Bao-An},
  journal = {Phys. Rev. C},
  volume = {76},
  issue = {5},
  pages = {054316},
  numpages = {25},
  year = {2007},
  month = {Nov},
  publisher = {American Physical Society},
  doi = {10.1103/PhysRevC.76.054316},
  url = {https://link.aps.org/doi/10.1103/PhysRevC.76.054316}
}

@article{sepah2003,
  title = {LOCV calculation for the uniform electron fluid at finite temperature},
  author = {Modarres, M and Moshfegh, H.R. and Sepahvand, A},
  journal = {Eur. Phys. J. B},
  volume = {31},
  pages = {159},
  year = {2003},
  doi = {10.1140/epjb/e2003-00020-0},

}

@article{Sammarruca_2010,
doi = {10.1088/0954-3899/37/8/085105},
url = {https://doi.org/10.1088/0954-3899/37/8/085105},
year = {2010},
month = {jun},
publisher = {},
volume = {37},
number = {8},
pages = {085105},
author = {Sammarruca, Francesca},
title = {Temperature dependence of single-particle properties in isospin-symmetric and -asymmetric matter within the Dirac–Brueckner–Hartree–Fock model},
journal = {Journal of Physics G: Nuclear and Particle Physics}
}

@article{RAMOS19891,
title = {Single-particle properties and short-range correlations in nuclear matter},
journal = {Nuclear Physics A},
volume = {503},
number = {1},
pages = {1-52},
year = {1989},
issn = {0375-9474},
doi = {https://doi.org/10.1016/0375-9474(89)90252-2},
url = {https://www.sciencedirect.com/science/article/pii/0375947489902522},
author = {A. Ramos and A. Polls and W.H. Dickhoff}
}

@article{PhysRevC.78.054003,
  title = {In-medium $T$ matrix for nuclear matter with three-body forces: Binding energy and single-particle properties},
  author = {Som\`a, V. and Bo\ifmmode \dot{z}\else \.{z}\fi{}ek, P.},
  journal = {Phys. Rev. C},
  volume = {78},
  issue = {5},
  pages = {054003},
  numpages = {9},
  year = {2008},
  month = {Nov},
  publisher = {American Physical Society},
  doi = {10.1103/PhysRevC.78.054003},
  url = {https://link.aps.org/doi/10.1103/PhysRevC.78.054003}
}

@article{PhysRevLett.90.152501,
  title = {Saturation of Nuclear Matter and Short-Range Correlations},
  author = {Dewulf, Y. and Dickhoff, W. H. and Van Neck, D. and Stoddard, E. R. and Waroquier, M.},
  journal = {Phys. Rev. Lett.},
  volume = {90},
  issue = {15},
  pages = {152501},
  numpages = {4},
  year = {2003},
  month = {Apr},
  publisher = {American Physical Society},
  doi = {10.1103/PhysRevLett.90.152501},
  url = {https://link.aps.org/doi/10.1103/PhysRevLett.90.152501}
}

@article{RIOS2007346,
title = {The entropy of a correlated system of nucleons},
journal = {Nuclear Physics A},
volume = {782},
number = {1},
pages = {346-349},
year = {2007},
note = {Proceedings of the 5th International Conference on Perspectives in Hadron Physics, Particle–Nucleus and Nucleus–Nucleus Scattering at Relativistic Energies},
issn = {0375-9474},
doi = {https://doi.org/10.1016/j.nuclphysa.2006.10.066},
url = {https://www.sciencedirect.com/science/article/pii/S0375947406007160},
author = {A. Rios and A. Polls and A. Ramos and H. Müther}
}

@article{FRIEDMAN1981205,
title = {The single particle potential in nuclear matter},
journal = {Physics Letters B},
volume = {100},
number = {3},
pages = {205-208},
year = {1981},
issn = {0370-2693},
doi = {https://doi.org/10.1016/0370-2693(81)90317-8},
url = {https://www.sciencedirect.com/science/article/pii/0370269381903178},
author = {B. Friedman and V.R. Pandharipande}
}

@article{MODARRES2023104047,
title = {The lowest order constrained variational (LOCV) method for the many-body problems and its applications},
journal = {Progress in Particle and Nuclear Physics},
volume = {131},
pages = {104047},
year = {2023},
issn = {0146-6410},
doi = {https://doi.org/10.1016/j.ppnp.2023.104047},
url = {https://www.sciencedirect.com/science/article/pii/S0146641023000285},
author = {Majid Modarres and Azar Tafrihi}
}

@article{OWEN1976170,
title = {Constrained variation in Jastrow method at high density},
journal = {Annals of Physics},
volume = {102},
number = {1},
pages = {170-188},
year = {1976},
issn = {0003-4916},
doi = {https://doi.org/10.1016/0003-4916(76)90260-8},
url = {https://www.sciencedirect.com/science/article/pii/0003491676902608},
author = {J.C Owen and R.F Bishop and J.M Irvine}
}

@article{HRMoshfegh_1998,
doi = {10.1088/0954-3899/24/4/012},
url = {https://doi.org/10.1088/0954-3899/24/4/012},
year = {1998},
month = {apr},
publisher = {},
volume = {24},
number = {4},
pages = {821},
author = {H R Moshfegh and M Modarres},
title = {The effect of three-body cluster energy on LOCV calculation for hot nuclear and neutron matter},
journal = {Journal of Physics G: Nuclear and Particle Physics}
}

@article{MOSHFEGH2007201,
title = {Thermal properties of asymmetrical nuclear matter with the new charge-dependent Reid potential},
journal = {Nuclear Physics A},
volume = {792},
number = {3},
pages = {201-218},
year = {2007},
issn = {0375-9474},
doi = {https://doi.org/10.1016/j.nuclphysa.2007.04.013},
url = {https://www.sciencedirect.com/science/article/pii/S0375947407004824},
author = {H.R. Moshfegh and M. Modarres}
}

@article{zar2010,
title = {A relativistic approach to the equation of state of asymmetric nuclear matter},
author = {Zaryouni, S. and Moshfegh, H. R.},
journal = {Eur. Phys. J. A},
volume = {45},
number = {1},
pages = {69-79},
year = {2010},
doi = {10.1140/epja/i2010-10983-1},
url = {https://link.springer.com/article/10.1140/epja/i2010-10983-1#citeas},

}

@article{SHAHRBAF201966,
title = {Appearance of hyperons in neutron stars within LOCV method},
journal = {Annals of Physics},
volume = {402},
pages = {66-77},
year = {2019},
issn = {0003-4916},
doi = {https://doi.org/10.1016/j.aop.2019.01.008},
url = {https://www.sciencedirect.com/science/article/pii/S0003491619300089},
author = {M. Shahrbaf and H.R. Moshfegh}
}

@article{PhysRevC.100.044314,
  title = {Equation of state and correlation functions of hypernuclear matter within the lowest order constrained variational method},
  author = {Shahrbaf, M. and Moshfegh, H. R. and Modarres, M.},
  journal = {Phys. Rev. C},
  volume = {100},
  issue = {4},
  pages = {044314},
  numpages = {12},
  year = {2019},
  month = {Oct},
  publisher = {American Physical Society},
  doi = {10.1103/PhysRevC.100.044314},
  url = {https://link.aps.org/doi/10.1103/PhysRevC.100.044314}
}

@article{Khanmo,
  title = {Hybrid star within the framework of a lowest-order constraint variational method},
  author = {Khanmohamadi S.,Moshfegh H.R.,  Atashbar Tehrani S.},
  journal = {Phys. Rev. D},
  volume = {101},
  issue = {2},
  pages = {023004},
  year = {2020},
  publisher = {American Physical Society},
  doi = {10.1103/PhysRevD.101.023004},
  url = {https://journals.aps.org/prd/abstract/10.1103/PhysRevD.101.023004}
}

@article{MODARRES20111,
title = {The single-particle potential of nuclear matter in the LOCV framework},
journal = {Nuclear Physics A},
volume = {867},
number = {1},
pages = {1-11},
year = {2011},
issn = {0375-9474},
doi = {https://doi.org/10.1016/j.nuclphysa.2011.07.009},
url = {https://www.sciencedirect.com/science/article/pii/S0375947411005628},
author = {M. Modarres and A. Rajabi}
}

@article{MOSHFEGH200579,
title = {Asymmetrical nuclear matter calculations with the new charge-dependent Reid potential},
journal = {Nuclear Physics A},
volume = {759},
number = {1},
pages = {79-91},
year = {2005},
issn = {0375-9474},
doi = {https://doi.org/10.1016/j.nuclphysa.2005.04.021},
url = {https://www.sciencedirect.com/science/article/pii/S037594740500610X},
author = {H.R. Moshfegh and M. Modarres}
}

@article{Goudarzi2015,
  author    = {S. Goudarzi and H. R. Moshfegh},
  title     = {Effects of three-body forces on the maximum mass of neutron stars in the lowest-order constrained variational formalism},
  journal   = {Physical Review C},
  volume    = {91},
  number    = {5},
  pages     = {054320},
  year      = {2015},
  doi       = {10.1103/PhysRevC.91.054320},
  publisher = {American Physical Society},
  url       = {https://doi.org/10.1103/PhysRevC.91.054320}
}

@article{bombaci1991asymmetric,
  title={Asymmetric nuclear matter equation of state},
  author={Bombaci, I and Lombardo, U},
  journal={Physical Review C},
  volume={44},
  number={4},
  pages={1892--1900},
  year={1991},
  publisher={American Physical Society},
  doi={10.1103/PhysRevC.44.1892},
  url={https://journals.aps.org/prc/abstract/10.1103/PhysRevC.44.1892}
}

@article{Wang_2020,
   title={Properties of nuclear matter in relativistic Brueckner–Hartree–Fock model with high-precision charge-dependent potentials},
   volume={47},
   ISSN={1361-6471},
   url={http://dx.doi.org/10.1088/1361-6471/aba423},
   DOI={10.1088/1361-6471/aba423},
   number={10},
   journal={Journal of Physics G: Nuclear and Particle Physics},
   publisher={IOP Publishing},
   author={Wang, Chencan and Hu, Jinniu and Zhang, Ying and Shen, Hong},
   year={2020},
  pages={105108} }

@article{Zuo_2012,
   title={Single particle potentials of asymmetric nuclear matter in different spin-isospin channels},
   volume={36},
   ISSN={1674-1137},
   url={http://dx.doi.org/10.1088/1674-1137/36/10/009},
   DOI={10.1088/1674-1137/36/10/009},
   number={10},
   journal={Chinese Physics C},
   publisher={IOP Publishing},
   author={Zuo, Wei and Gan, Sheng-Xin and Lombardo, U.},
   year={2012},
 pages={967–972} }

@article{PhysRevC.74.014317,
  title = {Three-body force rearrangement effect on single particle properties in neutron-rich nuclear matter},
  author = {Zuo, W. and Lombardo, U. and Schulze, H.-J. and Li, Z. H.},
  journal = {Phys. Rev. C},
  volume = {74},
  issue = {1},
  pages = {014317},
  numpages = {10},
  year = {2006},
  month = {Jul},
  publisher = {American Physical Society},
  doi = {10.1103/PhysRevC.74.014317},
  url = {https://link.aps.org/doi/10.1103/PhysRevC.74.014317}
}

@article{Dalen_2005,
   title={Momentum, density, and isospin dependence of symmetric and asymmetric nuclear matter properties},
   volume={72},
   ISSN={1089-490X},
   url={http://dx.doi.org/10.1103/PhysRevC.72.065803},
   DOI={10.1103/physrevc.72.065803},
   number={6},
   journal={Physical Review C},
   publisher={American Physical Society (APS)},
   author={Dalen, E. N. E. van and Fuchs, C. and Faessler, Amand},
   year={2005},
   month=Dec }

@article{Li2004,
  author = {Li, Bao-An},
  title = {Constraining the neutron-proton effective mass splitting in neutron-rich matter},
  journal = {Physical Review C},
  volume = {69},
  pages = {064602},
  year = {2004},
  month = {June},
  doi = {10.1103/PhysRevC.69.064602},
  issue = {6},
  publisher = {American Physical Society}
}

@article{LANE1962676,
title = {Isobaric spin dependence of the optical potential and quasi-elastic (p, n) reactions},
journal = {Nuclear Physics},
volume = {35},
pages = {676-685},
year = {1962},
issn = {0029-5582},
doi = {https://doi.org/10.1016/0029-5582(62)90153-0},
url = {https://www.sciencedirect.com/science/article/pii/0029558262901530},
author = {A.M. Lane}
}

@article{LiChen2005,
  author = {Li, Bao-An and Chen, Lie-Wen},
  title = {Nucleon-nucleon cross sections in neutron-rich matter and isospin transport in heavy-ion reactions at intermediate energies},
  journal = {Physical Review C},
  volume = {72},
  pages = {064611},
  year = {2005},
  month = {December},
  doi = {10.1103/PhysRevC.72.064611},
  issue = {6},
  publisher = {American Physical Society}
}

@article{Lane1962,
  author = {Lane, A. M.},
  title = {Isobaric spin dependence of the optical potential and quasi-elastic (p,n) reactions},
  journal = {Nuclear Physics},
  volume = {35},
  pages = {676-685},
  year = {1962},
  publisher = {North-Holland Publishing Co.},
  address = {Amsterdam}
}

@article{PhysRevC.82.054607,
  title = {Symmetry energy, its density slope, and neutron-proton effective mass splitting at normal density extracted from global nucleon optical potentials},
  author = {Xu, Chang and Li, Bao-An and Chen, Lie-Wen},
  journal = {Phys. Rev. C},
  volume = {82},
  issue = {5},
  pages = {054607},
  numpages = {5},
  year = {2010},
  month = {Nov},
  publisher = {American Physical Society},
  doi = {10.1103/PhysRevC.82.054607},
  url = {https://link.aps.org/doi/10.1103/PhysRevC.82.054607}
}

@article{PhysRevLett.94.032701,
  title = {Determination of the Stiffness of the Nuclear Symmetry Energy from Isospin Diffusion},
  author = {Chen, Lie-Wen and Ko, Che Ming and Li, Bao-An},
  journal = {Phys. Rev. Lett.},
  volume = {94},
  issue = {3},
  pages = {032701},
  numpages = {4},
  year = {2005},
  month = {Jan},
  publisher = {American Physical Society},
  doi = {10.1103/PhysRevLett.94.032701},
  url = {https://link.aps.org/doi/10.1103/PhysRevLett.94.032701}
}

@article{PhysRevC.72.064309,
  title = {Nuclear matter symmetry energy and the neutron skin thickness of heavy nuclei},
  author = {Chen, Lie-Wen and Ko, Che Ming and Li, Bao-An},
  journal = {Phys. Rev. C},
  volume = {72},
  issue = {6},
  pages = {064309},
  numpages = {5},
  year = {2005},
  month = {Dec},
  publisher = {American Physical Society},
  doi = {10.1103/PhysRevC.72.064309},
  url = {https://link.aps.org/doi/10.1103/PhysRevC.72.064309}
}

@article{LI2010509c,
title = {Recent Progress in Constraining the Equation of State of Dense Neutron-Rich Nuclear Matter with Heavy-Ion Reactions},
journal = {Nuclear Physics A},
volume = {834},
number = {1},
pages = {509c-514c},
year = {2010},
note = {The 10th International Conference on Nucleus-Nucleus Collisions (NN2009)},
issn = {0375-9474},
doi = {https://doi.org/10.1016/j.nuclphysa.2010.01.079},
url = {https://www.sciencedirect.com/science/article/pii/S0375947410000801},
author = {Bao-An Li and Lie-Wen Chen and De-Hua Wen and Zhigang Xiao and Chang Xu and Gao-Chan Yong and Ming Zhang}
}

@article{PhysRevC.81.064612,
  title = {Understanding the major uncertainties in the nuclear symmetry energy at suprasaturation densities},
  author = {Xu, Chang and Li, Bao-An},
  journal = {Phys. Rev. C},
  volume = {81},
  issue = {6},
  pages = {064612},
  numpages = {6},
  year = {2010},
  month = {Jun},
  publisher = {American Physical Society},
  doi = {10.1103/PhysRevC.81.064612},
  url = {https://link.aps.org/doi/10.1103/PhysRevC.81.064612}
}

@article{Li_2015,
   title={Neutron–proton effective mass splitting in neutron-rich matter at normal density from analyzing nucleon–nucleus scattering data within an isospin dependent optical model},
   volume={743},
   ISSN={0370-2693},
   url={http://dx.doi.org/10.1016/j.physletb.2015.03.005},
   DOI={10.1016/j.physletb.2015.03.005},
   journal={Physics Letters B},
   publisher={Elsevier BV},
   author={Li, Xiao-Hua and Guo, Wen-Jun and Li, Bao-An and Chen, Lie-Wen and Fattoyev, Farrukh J. and Newton, William G.},
   year={2015},
   month=Apr, pages={408–414} }

\end{document}